\documentclass[twocolumn]{aastex631}

\newcommand{\msun}{$M_{\odot}$}
\newcommand{\mstar}{$M_{\star}$}

\usepackage{hyperref}
\usepackage{amsmath}
\usepackage{booktabs}

\usepackage[table]{xcolor}
\definecolor{BrickRed}{RGB}{182,50,28}

\usepackage{array,booktabs,colortbl}
\newcolumntype{P}[1]{>{\raggedright\arraybackslash}p{#1}}

\begin{document}

\title{The quenched fraction of satellites around simulated Milky Way-mass galaxies}

\author[0000-0002-5908-737X]{Francisco J. Mercado}
\affiliation{Department of Physics \& Astronomy, Pomona College, Claremont, CA 91711, USA}
\email{francisco.mercado@pomona.edu}

\author[0000-0003-1848-5571]{Devontae C. Baxter}
\affiliation{Department of Astronomy \& Astrophysics, University of California, San Diego, 9500 Gilman Dr, La Jolla, CA 92093, USA}

\author[0000-0003-1848-5571]{M. Katy Rodriguez Wimberly}
\affiliation{Department of Physics \& Astronomy, California State University, San Bernardino, San Bernardino, CA 92407, USA}

\author[0000-0002-3430-3232]{Jorge Moreno}
\affiliation{Department of Physics \& Astronomy, Pomona College, Claremont, CA 91711, USA}
\affiliation{Carnegie Observatories, 813 Santa Barbara st, Pasadena, CA 91101, USA}

\author[0000-0002-2651-7281]{Coral Wheeler}
\affiliation{Department of Physics \& Astronomy, California Polytechnic State University Pomona, Pomona, CA 91768, USA}

\author[0000-0003-0965-605X]{Pratik Gandhi}
\affiliation{Department of Astronomy, Yale University, New Haven, CT 06520, USA}

\author[0000-0003-0603-8942]{Andrew Wetzel}
\affiliation{Department of Physics \& Astronomy, University of California, Davis, CA 95616, USA}

\author[0000-0002-1109-1919]{Robert Feldmann}
\affiliation{Department of Astrophysics, University of Zurich, Winterthurerstrasse 190, CH-8057 Zurich, Switzerland}

\author[0009-0005-8040-8325]{Lucas Tortora}
\affiliation{Institute of Astronomy and Kavli Institute for Cosmology, University of Cambridge, Madingley Road, Cambridge CB3 0HA, UK}

\author[0000-0002-8429-4100]{Jenna Samuel}
\affiliation{Department of Astronomy, The University of Texas at Austin,
Austin, TX 78712, USA}


\begin{abstract}
We compare satellite quenched fractions across three cosmological simulation suites (FIREbox, the FIRE-2 zoom-ins, and IllustrisTNG50) and observational datasets from SAGA, ELVES, and the combined satellite population of the Milky Way and M31. To enable consistent comparisons, we select Milky Way-mass hosts with $M_{\rm halo} = 10^{11.9}$ - $10^{12.2} \, M_{\odot}$ and satellites with stellar masses of $10^7$ - $10^{10} \, M_{\odot}$, applying nearly uniform radial selections and a common quenching definition. All three simulations reproduce the strong observed trend that lower-mass satellites are more likely to be quenched, closely matching the stellar mass dependence seen in SAGA, ELVES, and the Milky Way and M31 system. This agreement indicates that the mass dependence of satellite quenching is a robust outcome of contemporary galaxy formation models. Radial trends, however, show greater diversity. SAGA and ELVES exhibit gently declining quenched fractions with projected distance, consistent with stronger quenching at small radii. TNG50 most closely matches this behavior, while FIREbox remains broadly consistent with a weak radial trend within uncertainties. The FIRE-2 zoom-ins show suppressed quenched fractions at small projected distances, driven primarily by their paired MW-M31 analogs. We show that this discrepancy is not explained by host environment alone, but instead reflects atypical satellite populations in the paired systems, where star-forming and quenched satellites occupy distinct spatial distributions. Overall, our results demonstrate that stellar mass-quenched fraction trends are robust across simulations and observations, while radial trends are more sensitive to the detailed properties and distributions of satellite populations.
\end{abstract}



\section{Introduction}\label{sec:intro}  

Low-mass galaxies, with their shallow gravitational potentials, are highly susceptible to both \textit{in situ} and \textit{ex situ} processes that perturb the baryon cycle. This sensitivity makes them powerful probes of galaxy formation physics, including the galaxy–halo connection \citep{Nadler2020, Pina25}, the epoch of reionization \citep{Brown14, RodriguezWimberly19, McQuinn2024}, and the co-evolution of galaxies and their environments \citep{Weisz11, Geha12}. A key empirical result from environmental studies is the strong dichotomy between low-mass galaxies (\mstar\ $< 10^9$ \msun) that are isolated and those that are satellites of a more massive host. The former are almost exclusively star-forming \citep[with only a few rare exceptions;][]{Carleton24, Jiaxuan24, Giovanelli2013, Jones2024, Mutlu-Pakdil2025, McQuinn2024}, while the latter are far more likely to have ceased star formation (i.e., to be \textit{quenched}). 

Our understanding of how environment suppresses (or \textit{quenches}) star formation in low-mass galaxies is derived primarily from observations of the Local Group. Studies of the Millky Way and M31 satellite systems measure quenched fractions, inferred quenching timescales, and identified the dominant quenching mechanisms across nearly six decades in stellar mass \citep{Wheeler2014, Wetzel2015, RodriguezWimberly19, Fham2019}. Extending these analyses beyond the Local Group requires diverse observational strategies, including spectroscopic follow-up of faint satellites in nearby MW-like systems \citep{SAGA_I, SAGA_II, SAGA_IV}, volume-limited photometric surveys of nearby groups and Milky Way analogs \citep{Carlsten2021, Carlsten2022, ELVESIII, Meng2023}, as well as machine-learning and statistical background-subtraction techniques applied to Sloan Digital Sky Survey data \citep{Baxter2021}.

Despite methodological differences, these studies paint a consistent picture: below \mstar\ $\sim 10^{9.5}\,M_\odot,$ the satellite quenched fraction rises steadily toward lower stellar mass, reaching nearly 100\% for systems below \mstar\ $\sim 10^{5}\,M_\odot$ where Ultra-Faint Dwarfs are predominantly quenched by reionization. Moreover, the Satellites Around Galactic Analogs (SAGA) and the Exploration of the Local VolumE (ELVES) surveys find that quenched fractions around MW-mass hosts remain low ($\sim$0.2) at most host-centric radii within 300 kpc, with significant increases only in the innermost regions \citep{ELVESIII, SAGA_IV}. These results provide a robust observational framework against which theoretical models can be tested.

A number of recent studies explore satellite quenching around MW-like hosts in cosmological simulations, consistently finding qualitative agreement with the observed mass-dependent trend \citep{Simpson2018, Akins2021, Karunakaran2021, Font2022, Samuel2022, Engler2023, Christensen2024, Rodriguez-Cardoso2025}. This agreement holds across diverse physics implementations, numerical resolutions, and choices of quenching diagnostics. For example, \citet{Samuel2022} examine quenching in the FIRE-2 MW-mass zoom-in simulations using time-resolved star formation histories and HI gas content, while other studies employ proxies such as youngest star-particle age or instantaneous specific star formation rate (sSFR) \citep[e.g.,][]{Simpson2018, Font2022, Engler2023}. Time-averaged sSFR measurements tied to observational tracers (H$\alpha$, ultraviolet and far-ultraviolet) also yield similar qualitative trends \citep{Akins2021, Karunakaran2021}. Cross-code comparisons from the Assembling Galaxies Of Resolved Anatomy project \citep[AGORA;][]{AGORA2014} show that this trend persists even when identical initial conditions are evolved with different hydrodynamic solvers \citep{Rodriguez-Cardoso2025}.  \citet{Engler2023} likewise demonstrate this behavior in the IllustrisTNG50 volume simulation \citep{Nelson2018, Pillepich2018}, although the normalization of quenched fractions depends sensitively on how MW-like hosts are selected. Notably, the SAGA collaboration showed that the same rising quenched fraction toward lower stellar masses emerges in the UNIVERSEMACHINE empirical framework \citep{UniverseMachine, SAGA_V}, which forward models galaxy star formation histories based on halo assembly rather than hydrodynamics.

While the mass-dependent trend appears robust, considerably less attention has been devoted to whether simulations reproduce the observed radial dependence of quenching within $\sim1 \,r_{\rm vir}$. This is a key gap: environmental processes such as ram-pressure stripping, starvation, and tidal interactions are strongest near the host and should imprint clear signatures on radial quenched fraction profiles. Understanding the degree to which simulations capture this behavior is essential for disentangling the roles of internal physics and environment in satellite quenching.

Comparing satellite populations across simulations is inherently challenging: different halo-finding algorithms, feedback models, resolutions, and stellar-mass definitions mean that host halos selected to have the same mass are not strictly equivalent across suites. As a result, truly one-to-one comparisons require a level of homogenization beyond the scope of most existing studies. Here we instead take a first-order approach, showing that even under these mismatched but internally consistent definitions, the qualitative mass-dependent trend in quenched fractions remains robust.

In this paper, we examine satellite quenching as a function of both stellar mass and projected host-centric distance for Milky Way–mass galaxies across three simulation suites: FIREbox \citep{Feldmann2023}, the FIRE-2 MW-mass zoom-ins \citep{Wetzel2016, GK2017, GK2019a, GK2019b, Samuel2020}, and IllustrisTNG50 \citep{Nelson2018, Pillepich2018}. We compare these predictions directly to satellite populations observed in the SAGA Survey \citep{SAGA_IV}, the ELVES Survey \citep{ELVESIII}, and the combined MW–M31 system \citep{Wetzel2016}. Despite substantial differences in resolution, cosmological volume, feedback implementations, and halo-finding architectures, all three simulations reproduce the observed mass dependence of satellite quenching. Their radial quenched fraction trends, however, show broader diversity. The largest discrepancy appears in the paired FIRE-2 systems, where quenched fractions are suppressed at small projected radii. We show that this behavior is associated with an atypical satellite population in the paired FIRE-2 hosts, characterized by a distinct combination of satellite stellar masses, radii, and quenching states, rather than with a universal effect of host environment.

The remainder of this paper is organized as follows. Section~\ref{sec:methodology} describes the simulations, observational datasets, host and satellite selection criteria, and methods for computing stellar and star formation properties. Section~\ref{sec:results} presents satellite population statistics and global quenched fraction trends across both simulations and observations. We also investigate the role of larger-scale host environment in shaping quenched fractions and examine the discrepant radial behavior exhibited by the FIRE-2 zoom-in simulations. Section~\ref{sec:discussion} interprets these results in the context of feedback physics, environmental processes, and host-galaxy properties. Section~\ref{sec:conclusion} summarizes our main findings and highlights future work, including two forthcoming studies that will quantify how host environment and host properties jointly regulate satellite quenching.

\section{Methodology}\label{sec:methodology}

\subsection{The FIRE-2 Galaxy Formation Model}\label{sec:FIRE2_model}

All FIRE simulations used in this work, including the cosmological volume FIREbox run and the FIRE-2 zoom-in suite, employ the FIRE-2 galaxy formation model \citep{Hopkins2018}. The simulations are run with the \textsc{GIZMO} code \citep{Hopkins2015} using the meshless finite-mass (MFM) hydrodynamics method. The FIRE-2 model includes explicit treatments of radiative cooling, star formation, and stellar feedback, and resolves the multiphase interstellar medium at sub-kiloparsec scales.

Gas cools radiatively to below $10^4$\,K and forms stars only in dense, self-gravitating, molecular, and self-shielding gas. FIREbox adopts a star-formation threshold of $n = 300\,{\rm cm^{-3}}$, and the zoom-in simulations use a threshold of $n = 1000\,{\rm cm^{-3}}$. Stellar feedback includes momentum, energy, mass, and metal return from Type Ia and core-collapse supernovae, winds from massive stars and asymptotic giant branch stars, radiation pressure, and both photo-electric and photo-ionization heating, following STARBURST99 stellar population models \citep{Leitherer99}. Metal enrichment is tracked for 11 species, and the simulations assume a Kroupa IMF \citep{Kroupa02}. A spatially uniform and redshift-dependent UV and X-ray background \citep{CAFG2009} is applied.

Initial conditions for FIRE simulations are generated with \textsc{MUSIC} \citep{Hahn2011} using a flat $\Lambda$CDM cosmology consistent with \citet{Planck2016}. Although FIREbox and the zoom-in simulations differ in volume and resolution, they share identical baryonic physics. This allows direct comparisons across environments and host masses without introducing differences in subgrid prescriptions.

\subsection{The FIREbox Cosmological Volume Simulation}\label{sec:FIREbox}

FIREbox \citep{Feldmann2023} is a periodic cosmological volume with a side length of $22.1\,{\rm cMpc}$, evolved from $z=120$ to $z=0$ with $1024^3$ gas and $1024^3$ dark matter particles. The flagship \texttt{FB1024} run in this work has baryonic and dark matter particle masses of $m_{\rm b}=6.3\times10^4\,M_\odot$ and $m_{\rm DM}=3.3\times10^5\,M_\odot$, and adopts gas softening that is adaptive down to 1.5 pc. Star and dark matter particles have fixed softenings of 12\,pc and 80\,pc, respectively.

Although FIREbox is comparatively more modest in volume than some large cosmological simulations \citep[e.g.,][]{Schaye2015, Pillepich2018}, it resolves scales from a few parsecs to several megaparsecs and includes environments ranging from voids to small galaxy groups with halo masses approaching $10^{13}\,M_\odot$. The simulation’s high spatial resolution enables realistic modeling of interstellar physics, making it ideal for studying low-mass galaxy structure and environmental effects \citep{Mercado2025,Klein2025,Benavides2025,Moreno2025}.

\subsubsection{Halo identification in FIREbox}\label{sec:FIREbox_halo_identification}

We identify halos in FIREbox with the AMIGA Halo Finder (AHF; \citealt{Knollmann2009}), which reports virial quantities using a redshift-dependent overdensity threshold relative to the mean matter density \citep{Bryan1998}. To place FIREbox on the same halo definition used for the FIRE-2 zoom-in simulations and TNG50, we convert these virial quantities to a spherical overdensity definition based on the mean matter density. We adopt the \citet{Bullock2001} mass–concentration relation and solve for the radius and enclosed mass corresponding to an average density of 200 times the mean matter density, yielding
\begin{equation}
r_{\rm halo} = r_{200{\rm m}}, \qquad M_{\rm halo} = M_{200{\rm m}}.
\end{equation}

For satellite subhalos, AHF reports a truncation radius that encloses the gravitationally bound region. Because our analysis does not make use of satellite halo masses, we use only the AHF-provided positions and $r_{\rm t}$ values to identify bound satellites and compute their distances from host galaxies.

We adopt these halo radius and mass definitions for all simulation-based analyses throughout this work, ensuring consistency across FIREbox, the FIRE-2 zoom-in simulations, and TNG50.

\subsection{FIRE-2 Zoom-in Simulation Suite}\label{sec:FIREzooms}

To complement the FIREbox volume, we analyze 14 high-resolution FIRE-2 zoom-in simulations of Milky Way or Andromeda mass galaxies at $z=0$. These systems span a range of environments:
\begin{itemize}
    \item Eight isolated systems from the Latte suite \citep{Wetzel2016, GK2017, Samuel2020}, with $M_{200{\rm m}}\sim1$ to $2\times10^{12}\,M_\odot$ and no comparably massive neighbors within $5R_{200{\rm m}}$.
    \item Six paired systems from the ELVIS on FIRE suite \citep{GK2017, GK2019b}, selected to reproduce Local Group like conditions with halo separations of 600 to 1000 kpc.
\end{itemize}

The isolated systems use dark matter particle masses of $3.5\times10^4\,M_\odot$ and initial gas or star particle masses of approximately $7\times10^3\,M_\odot$, with gas softening as small as 1 pc. The paired systems have roughly twice the mass resolution. Previous FIRE resolution studies show that such differences lead to only modest variations in galaxy properties and do not affect quenched fraction measurements in any significant way \citep{Hopkins2018}.

The FIRE-2 zoom-in simulations reproduce many observed features of the Local Group. These include stellar mass assembly histories, metallicity distributions, disk morphologies, and the structure of their satellite populations \citep{Ma2017, Sanderson2020, Samuel2022, McCluskey2024}. Their satellite stellar mass functions, radial distributions, and quenching timescales broadly agree with those observed for the Milky Way and M31, particularly below the LMC/SMC mass scale \citep{Wetzel2016, GK2019a, Samuel2020}. We note, however, that the FIRE-2 hosts were not selected to reproduce the specific satellite populations of the Milky Way or M31, and several systems lack satellites as massive as the LMC or M33. Such massive companions are statistically uncommon around Milky Way–mass halos \citep{Busha2011, Tollerud2011}. These comparisons nevertheless provide confidence that the FIRE-2 MW-mass systems serve as realistic analogs of Local Group environments over the satellite stellar-mass range most relevant to this study.

\subsubsection{Halo identification in the FIRE-2 zooms}\label{sec:FIREzooms_halo_identification}

Halos and subhalos in the FIRE-2 zoom-in simulations are identified using the \textsc{ROCKSTAR} halo finder \citep{Behroozi2013}, which performs adaptive refinement in phase space and uses temporal information to track halo evolution. ROCKSTAR directly reports spherical overdensity quantities relative to the mean matter density, which match the unified halo definition adopted for this work.

Because we apply a single halo-mass and halo-radius definition across all simulations, we do not recalculate or modify ROCKSTAR-reported quantities for the FIRE-2 zooms, nor do we attempt to re-derive FIREbox halos under ROCKSTAR conventions. This approach preserves the physically meaningful differences between halo finders while ensuring consistent host selection and distance measurements.

\subsection{IllustrisTNG: TNG50}\label{sec:TNG50}

We also include Milky Way–mass hosts from the TNG50 simulation \citep{Nelson18, Pillepich18, Springel18}, the highest-resolution run in the IllustrisTNG project. TNG50 evolves a $(51.7\, {\rm cMpc})^3$ cosmological volume with the moving-mesh \textsc{AREPO} code \citep{Springel10}, which solves ideal magnetohydrodynamics using a finite-volume Godunov scheme on a moving Voronoi mesh. The galaxy formation model includes metal-line cooling, star formation in a pressurized interstellar medium, kinetic and thermal energy injection from supernovae, and both thermal and kinetic modes of black hole feedback \citep{Weinberger2017}. The \texttt{TNG50-1} run used in this work reaches baryonic and dark matter mass resolutions of $8.5\times10^4\,M_\odot$ and $4.5\times10^5\,M_\odot$, with gravitational softening lengths of approximately 300 pc for collisionless particles.

We use TNG50 rather than the larger TNG100 or TNG300 volumes because its mass and spatial resolution are sufficient to robustly resolve the low-mass satellite population relevant for this study. In particular, TNG50 reaches baryonic and dark matter mass resolutions of $8.5\times10^4\,M_\odot$ and $4.5\times10^5\,M_\odot$, which allows us to model satellites down to the stellar masses probed by SAGA, ELVES, and the FIRE simulations. This makes TNG50 an appropriate comparison set for examining satellite quenching at the low-mass scales of interest.

\subsubsection{Halo identification in TNG50}\label{sec:TNG50_halo_identification}

Halos in TNG50 are first identified using a Friends-of-Friends (FoF) algorithm \citep{Davis1985}, after which the \textsc{SUBFIND} substructure finder \citep{Springel2001} locates gravitationally bound subhalos within each FoF group. For central galaxies, we use the spherical overdensity quantities reported by TNG50 and adopt their $200$-times-the-mean–density values as the halo radii and masses for host selection, consistent with the unified definition applied throughout this work.

Satellite subhalos are treated as bound structures within the FoF group. Because \textsc{SUBFIND} does not assign spherical-overdensity radii (e.g.\ $r_{200\mathrm{m}}$) to satellites, we rely only on their positions, stellar masses, and star formation rates when computing distances from the host and evaluating satellite quenching. No subhalo radius is required for our analysis.

\begin{table*}[t]
\centering
\caption{
Summary of the numerical characteristics, halo definitions, and feedback implementations in the three simulation suites analyzed in this work. Each suite retains its halo-finding conventions and stellar-mass definitions. The table lists the dark-matter and baryonic particle masses, spatial resolutions, feedback physics, and the halo and stellar-mass conventions used in FIREbox, the FIRE-2 zoom-ins, and TNG50. These simulation-specific quantities define the physical and numerical context for our comparison to observational datasets.
}
\label{tab:summary}

\small
\renewcommand{\arraystretch}{1.25}

\begin{tabular*}{\textwidth}{@{\extracolsep{\fill}} llll}
\hline\hline
Property & FIREbox & FIRE-2 zooms & TNG50 \\
\hline\hline

\multicolumn{4}{l}{\textbf{Mass resolution, spatial resolution, and feedback}} \\
\hline
\addlinespace[4pt]

DM particle mass &
$3.3\times10^{5}\,M_\odot$ &
\shortstack[l]{Isolated: $3.5\times10^{4}\,M_\odot$\\Paired: $1.9$--$2.0\times10^{4}\,M_\odot$} &
$4.5\times10^{5}\,M_\odot$ \\
\addlinespace[4pt]

Baryon particle mass &
$6.3\times10^{4}\,M_\odot$ &
\shortstack[l]{Isolated: $7.1\times10^{3}\,M_\odot$\\Paired: $3.5$--$4.0\times10^{3}\,M_\odot$} &
$8.5\times10^{4}\,M_\odot$ \\
\addlinespace[4pt]

Spatial resolution &
\shortstack[l]{Stars: 12 pc\\Gas: adaptive to 1.5 pc} &
\shortstack[l]{Stars: 4 pc\\Gas: adaptive to 1 pc} &
\shortstack[l]{Collisionless: 290 pc\\Gas: adaptive to 74 pc} \\
\addlinespace[4pt]

Feedback channels &
\shortstack[l]{SNe Ia/II; stellar winds;\\photo-ionization; radiation pressure} &
Same as FIREbox &
\shortstack[l]{SNe winds; AGN\\thermal + kinetic feedback} \\
\addlinespace[6pt]

\hline
\multicolumn{4}{l}{\textbf{Halo definitions}} \\
\hline
\addlinespace[4pt]

Halo finder &
AHF & ROCKSTAR & SUBFIND \\
\addlinespace[4pt]

$M_{\rm halo}$ definition &
$M_{\rm vir}\rightarrow M_{200{\rm m}}$ &
$M_{200{\rm m}}$ &
$M_{200{\rm m}}$ \\
\addlinespace[4pt]

$r_{\rm halo}$ definition &
$r_{200{\rm m}}$ &
$r_{200{\rm m}}$ &
$r_{200{\rm m}}$ \\
\addlinespace[4pt]

Stellar-mass definition &
Bound stars within $r^{\star}_{80}$ &
Bound stars within $r^{\star}_{80}$ &
Total bound stellar mass \\
\addlinespace[4pt]

\hline\hline
\end{tabular*}
\end{table*}

A summary of the numerical characteristics, halo-finding conventions, and stellar-mass definitions used for simulations is provided in Table~\ref{tab:summary}.

\subsection{Observational datasets}
\label{sec:obs_data}

To complement the simulation suites, we incorporate two large observational surveys of Milky Way--mass galaxies: the SAGA and ELVES surveys. Both provide uniform satellite catalogs around statistically significant samples of nearby MW-mass hosts, enabling a direct comparison between synthetic and observed satellite populations in the low-mass regime relevant for quenching studies.

\subsubsection{SAGA: Satellites Around Galactic Analogs}\label{sec:SAGA}
We use the public Data Release 3 (DR3) from the SAGA Survey\footnote{\url{https://sagasurvey.org/data/}}, which provides photometric and spectroscopic measurements for satellites around 100 Milky Way analogs out to projected distances of 300 kpc \citep[see][]{SAGA_III,SAGA_IV}. Our analysis focuses on the confirmed and high-probability (``Gold + Silver'') satellite samples, corresponding to the DR3 \texttt{sample < 3} selection. Host halo masses in SAGA are estimated following the group catalog of \citet{Lim2017}, which assigns $M_{\rm halo}$ via abundance matching and group-finding applied to SDSS. \emph{These halo-mass estimates are subject to substantial uncertainties inherent to group-finding and abundance-matching techniques. As a result, comparisons between SAGA halo masses and those from simulations should be interpreted with caution.} For each satellite, DR3 reports stellar masses, H$\alpha$ equivalent widths, projected separations, and catalog-level quenched classifications. We reconstruct the quenched flag using the DR3 spectroscopic criterion to ensure consistency with the definition used in the SAGA collaboration papers (see Section~\ref{sec:quenched_definition}).

\subsubsection{ELVES: Exploration of the Local VolumE Satellites}\label{sec:ELVES}
We also use the compiled ELVES \citep{ELVESIII} satellite catalog released with SAGA Paper IV\footnote{\url{https://github.com/sagasurvey/saga-paper4}} \citep{SAGA_IV}, which includes confirmed and candidate satellites around 28 nearby galaxies with K-band luminosities similar to that of the Milky Way. The catalog provides homogenized photometry, distances, stellar masses, quenched classifications, and projected separations for all satellites. Unlike SAGA, this ELVES catalog does not report halo mass estimates for its hosts; we therefore employ ELVES only for satellite-level comparisons and for analyses that do not require host halo masses.

Together, SAGA and ELVES provide complementary observational constraints: SAGA offers a large, statistically uniform sample with probabilistic satellite identification, while ELVES offers deeper coverage of the nearest systems with high-quality stellar population measurements. These datasets serve as key empirical benchmarks for evaluating the quenched fraction and structural properties of satellites in our simulated host samples.

\subsection{Stellar-mass definitions}
\label{sec:stellar_mass_definition}

Because stellar-mass measurements differ across simulations and surveys, we adopt dataset-specific definitions that each capture the total bound stellar mass of a satellite, ensuring physical consistency across all datasets.

For both FIREbox and the FIRE-2 zoom-in simulations, we define the stellar mass of each galaxy as the total mass of all \emph{gravitationally bound} star particles enclosed within the radius $r^{\star}_{80}$, which contains 80\% of the bound stellar mass. This choice minimizes sensitivity to tidal debris, optimizes consistency across halo finders (AHF for FIREbox and ROCKSTAR for FIRE-2), and provides a robust measure of the stellar component for low-mass satellites.

For TNG50, we adopt the SUBFIND-reported stellar mass, which corresponds to the total gravitationally bound stellar mass of the subhalo. Because SUBFIND does not assign spherical-overdensity radii to satellites, we use the cataloged bound stellar mass directly without imposing an additional aperture.

For both observational samples, we take the stellar masses provided in the publicly released catalogs. SAGA DR3 stellar masses are derived from optical+UV photometry using SED fitting (see \citealt{SAGA_IV}). ELVES stellar masses come from homogeneous optical photometry calibrated to deep imaging of the Local Volume \citep{ELVESIII}. We apply no additional aperture corrections or rescaling.

These definitions are intended to approximate the bound stellar components of satellites across all datasets while respecting the conventions and resolution limits of each simulation suite and survey. We caution, however, that systematic differences in stellar-mass definitions between simulations and observations may shift satellite stellar masses by of order one to two tenths of a dex, potentially moving individual satellites between adjacent stellar-mass bins. We have verified that shifts of this magnitude do not qualitatively affect the trends presented in this work, although they can introduce modest variations in individual bins.

\subsection{Host selection}
\label{sec:host_selection}

To enable consistent comparisons across all datasets, we identify Milky Way–mass hosts using a halo mass range of
\begin{equation}
10^{11.9} \leq M_{\rm halo}/M_\odot \leq 10^{12.2}.
\end{equation}
This interval encompasses the range of recent Milky Way halo-mass estimates derived from stellar kinematics, satellite dynamics, and mass modeling \citep[e.g.,][]{Bland-Hawthorn2016, Watkins2010, Patel2017}. It also covers the target masses of the FIRE-2 zoom-in simulations, ensuring that all simulated hosts can be treated within a common selection framework.

For FIREbox, the FIRE-2 zooms, and TNG50, we first apply this halo-mass cut to all central galaxies. We then construct parent neighbor catalogs for each host, including all galaxies within 1~Mpc. We apply the satellite stellar-mass selection described in Section~\ref{sec:satellite_selection} to these neighbor catalogs prior to constructing projected mock observations. Hosts with no neighboring galaxies in this stellar-mass range are removed from the final host sample.\footnote{This results in only one host from FIREbox being removed from the sample.} After all selections, the resulting numbers of Milky Way–mass hosts are 19 for FIREbox, 14 for the FIRE-2 zooms, and 140 for TNG50.

For the SAGA survey, we retain all hosts to preserve statistical power and avoid uncertainties in the cataloged halo-mass estimates. Applying the same halo-mass and satellite cuts used for the simulations would reduce the SAGA sample to 46 hosts, but the corresponding quenched fractions and radial trends are fully consistent with those obtained from the full set. After applying the stellar-mass completeness requirement to SAGA satellites, the final SAGA sample consists of 97 hosts.

For ELVES, we impose two observational cuts. First, we apply a simple absolute-$K$-band selection,
\begin{equation}
-24.6 \leq M_{K} \leq -23.0,
\end{equation}
which approximates the SAGA host-luminosity range and serves as a coarse proxy for Milky Way–mass systems. Second, because many ELVES hosts have imaging and spectroscopy only out to 150--200 kpc, we additionally require a radial coverage of at least 300 kpc, using the cataloged \texttt{r\_cover} values. This ensures that ELVES hosts are selected on a comparable projected aperture to that used for the simulated satellite samples. After applying both the $K$-band and radial-coverage cuts, and enforcing satellite stellar-mass completeness, our final ELVES sample contains 14 hosts.

\begin{figure*}
\centering
\includegraphics[width=\textwidth]{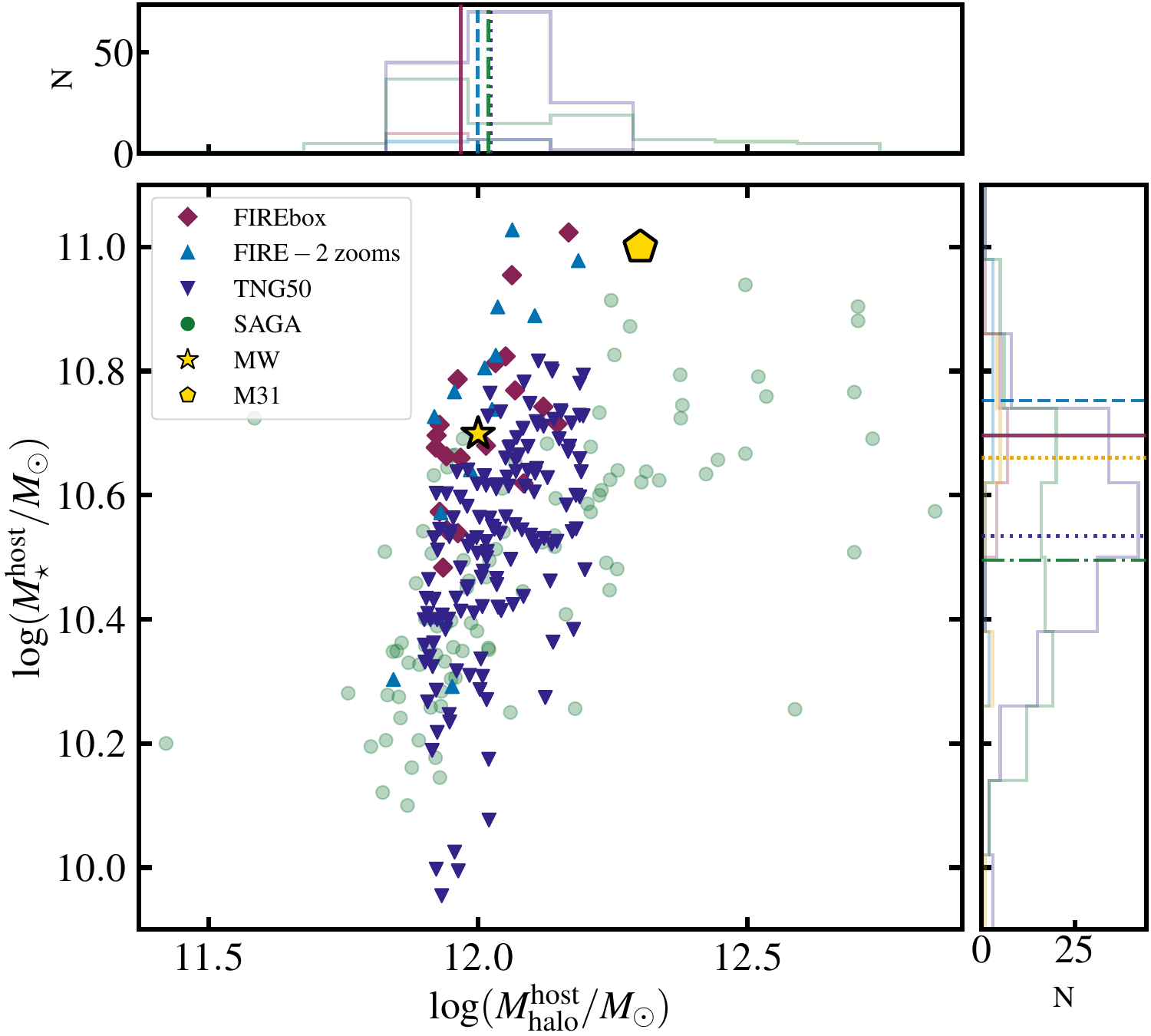}
\caption{
Stellar mass–halo mass (SMHM) relation for host samples drawn from FIREbox (maroon), the FIRE-2 zoom-ins (blue), TNG50 (indigo), and the SAGA survey (green). All simulated hosts lie within the adopted Milky Way–mass range of $10^{11.9} \leq M_{\rm halo}/M_\odot \leq 10^{12.2}$, while the full SAGA sample is retained to preserve statistical completeness and account for uncertainties in cataloged halo masses. The Milky Way and M31 are shown as a yellow star and pentagon. Histograms of host halo mass (top) and stellar mass (right) share axes with the main panel. The median halo masses are $\log(M_{\rm halo}/M_\odot) = 11.97$ (FIREbox), 12.00 (FIRE-2 zooms), 12.02 (TNG50), and 12.02 (SAGA). The median stellar masses are $\log(M_\star/M_\odot)=10.70$ (FIREbox), 10.75 (FIRE-2 zooms), 10.53 (TNG50), 10.50 (SAGA), and 10.66 (ELVES). Because ELVES does not provide halo-mass estimates, its hosts appear only in the stellar-mass histogram.
}
\label{fig:smhm_hosts}
\end{figure*}

Figure~\ref{fig:smhm_hosts} summarizes the stellar mass–halo mass (SMHM) distributions of the resulting host populations. All simulated hosts lie within the adopted halo-mass interval, while the full SAGA sample is retained to reflect uncertainties in observational halo-mass estimates. The Milky Way and M31 are overplotted for reference.

Across all datasets, the median halo masses are remarkably consistent, with $\log(M_{\rm halo}/M_\odot) \approx 12.0$ for FIREbox (11.97), the FIRE-2 zooms (12.00), TNG50 (12.02), and SAGA (12.02). The corresponding stellar masses show modest variation: FIREbox and the FIRE-2 zooms have median stellar masses of $\log(M_\star/M_\odot) = 10.70$ and 10.75, respectively, while TNG50 and SAGA hosts are slightly lower at 10.53 and 10.50. The ELVES hosts have a median stellar mass of 10.66 and appear only in the stellar-mass distribution due to the absence of halo-mass estimates.

\begin{figure}
    \centering
    \includegraphics[width=\linewidth]{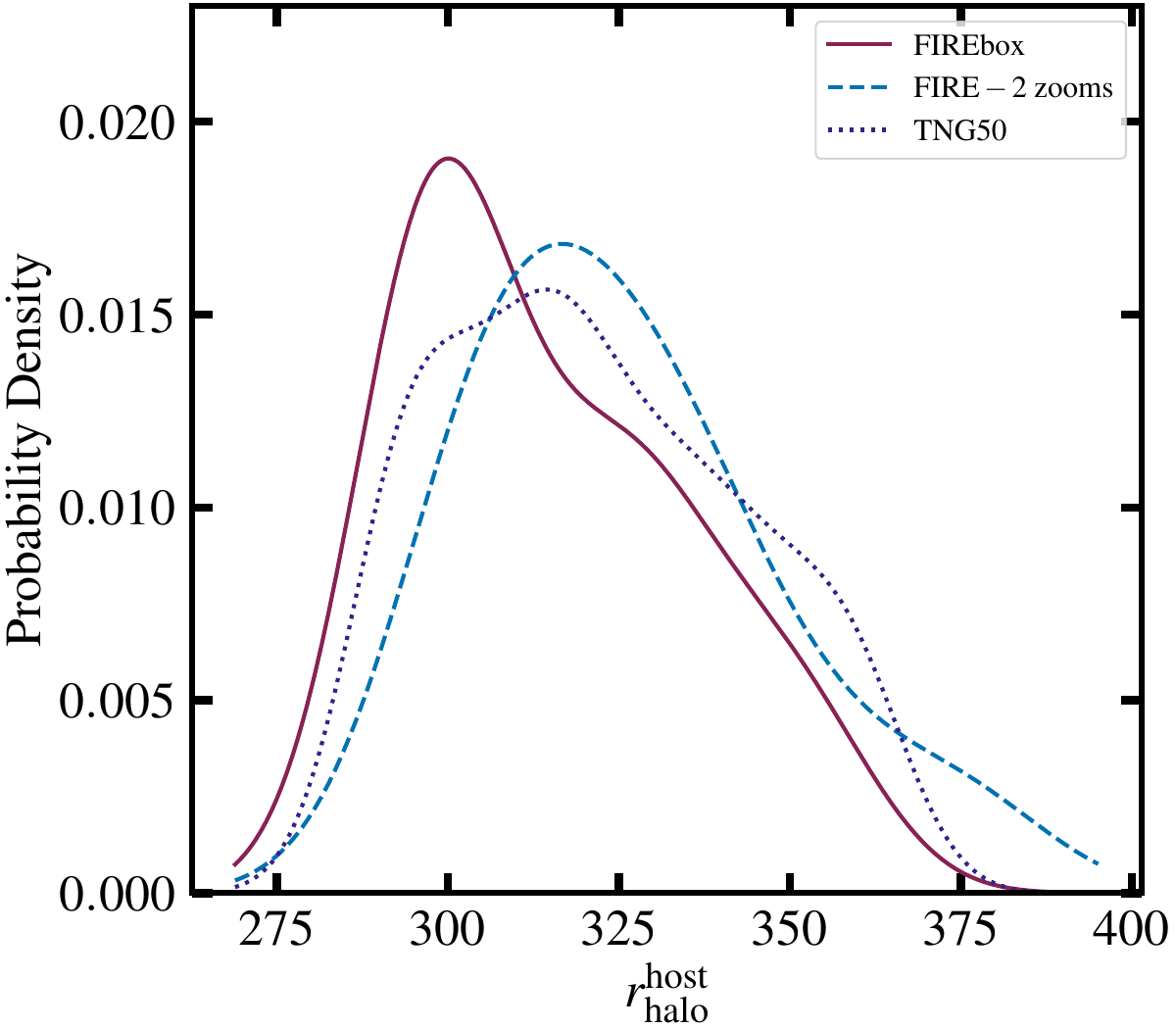}
    \caption{Kernel density estimates of host halo radii, $r_{\rm halo}$, for Milky Way--mass systems drawn from FIREbox (maroon), the FIRE-2 zoom-ins (blue), and TNG50 (indigo). The distributions agree closely, with median halo radii of 304.6 kpc, 318.8 kpc, and 317.8 kpc for FIREbox, the FIRE-2 zoom-ins, and TNG50, respectively. The modest differences between the distributions reflect variations in halo-finding methodology and halo definitions across the simulation suites.}
    \label{fig:halo_radius_kde}
\end{figure}

Although the simulations target comparable halo masses, differences in halo-finding methodology introduce modest systematic variations in the recovered halo radii. In Figure~\ref{fig:halo_radius_kde} we show kernel density estimates of the host halo radii for each simulation suite. The distributions agree closely, with median halo radii of 304.6 kpc for FIREbox, 318.8 kpc for the FIRE-2 zoom-ins, and 317.8 kpc for TNG50. These relatively small offsets support our use of fixed physical apertures when constructing projected satellite catalogs, while also highlighting that modest differences in halo definitions can contribute to variations in satellite selection and interloper fractions across simulations.

These similarities demonstrate that our host selection produces broadly comparable Milky Way–mass populations across all datasets, enabling meaningful comparisons of satellite populations.

\subsection{Satellite selection}
\label{sec:satellite_selection}

We select satellites across all datasets using criteria matched to the mass-resolution limits and observational completeness thresholds of the SAGA and ELVES surveys. For the simulations, we begin with parent neighbor catalogs that include all galaxies within 1 Mpc of each host. We first apply a stellar-mass selection to define the satellite population in three dimensions prior to constructing projected mock catalogs. Because line-of-sight satellite positions in the SAGA dataset are uncertain at the level of several hundred kiloparsecs, we then mimic this effect by projecting each host--satellite system along the XY, XZ, and YZ axes of the simulation box and selecting satellites within a cylindrical aperture centered on the host.

Within the parent neighbor catalogs, we require satellite stellar masses to lie within
\begin{equation}
7 \leq \log_{10}(M_{\star}/M_\odot) \leq 10.
\end{equation}
The lower bound reflects the stellar-mass completeness imposed by the coarsest resolved simulation in our comparison. TNG50 resolves satellites with $M_\star \approx 10^{7}\,M_\odot$ using more than one hundred stellar particles, ensuring that systems above this threshold are well resolved across all three simulation suites. The upper bound excludes unusually massive companions whose stellar masses approach those of Milky Way–mass systems, allowing us to compare satellite populations across simulations and observations using a consistent and conservative definition.

After applying this stellar-mass selection, we construct mock observational catalogs by selecting satellites that satisfy
\begin{equation}
10 \, {\rm kpc} \leq d_{\rm proj} \leq 300 \,{\rm kpc}
\end{equation}
and
\begin{equation}
|d_{\rm los}| \leq 500\,{\rm kpc},
\end{equation}
where $d_{\rm proj}$ is the projected host-centric separation and $d_{\rm los}$ is the line-of-sight separation in the chosen projection. This corresponds to a cylindrical aperture of projected radius 300 kpc and total depth of 1 Mpc, matching the scale of the line-of-sight uncertainties in SAGA. The inner 10 kpc are excluded to match the observational selection used in \citet{SAGA_IV}, where host-galaxy light strongly suppresses satellite detectability.

These same stellar-mass limits are applied to the observed satellites in both SAGA and ELVES to ensure that all five datasets are compared over a common satellite mass range. Because each simulated system is viewed along three orthogonal sightlines, the mock catalogs naturally include projection-driven interlopers, just as the observational samples do. We retain these interlopers by construction so that the simulations and surveys are compared on the same observational footing.

\subsection{Host environment classification}
\label{sec:environment_definition}

We classify host galaxy environments using criteria designed to closely follow those adopted in the SAGA Survey. In the SAGA public data release, we use the \texttt{sep-massive} quantity, which provides the distance to the nearest galaxy with stellar mass comparable to or greater than the host, including Milky Way--mass systems. We define a host as \emph{isolated} if no such massive companion exists within 1 Mpc (i.e., $\texttt{sep-massive} > 1$ Mpc), and as \emph{non-isolated} otherwise.

To enable a consistent comparison, we apply an analogous definition to the simulations. For each Milky Way--mass host, we identify all central galaxies with halo masses comparable to or greater than the lower limit of the host sample ($M_{\rm halo} \geq 10^{11.9}\,M_\odot$) and compute their three-dimensional distances from the host. A host is classified as isolated if no such massive companion lies within 1 Mpc, and as non-isolated otherwise.

While the datasets allow for more detailed environment classifications (e.g., distinguishing Local Group-like pairs from systems with multiple massive companions), we do not separate these cases in the main analysis. Instead, we adopt a binary division between isolated and non-isolated hosts to maintain consistency with the observational data and ensure sufficient sample sizes across all datasets.

This definition provides a simple and physically motivated characterization of environment, enabling a direct comparison of satellite populations across simulations and observations.

\subsection{Quenched classification}\label{sec:quenched_definition}

Observational quenching diagnostics rely on tracers of recent star formation such as H$\alpha$, UV/FUV continuum, and cold-gas emission \citep{Leroy2008, Grcevich2009, Spekkens2014, Putman2021}. These indicators probe different characteristic timescales, with H$\alpha$ tracing star formation over approximately 10 Myr and UV/FUV tracing activity over roughly 100 Myr. Cosmological simulations provide access to the full star formation histories of individual star particles, which allows both time-resolved and observation-matched quiescence definitions \citep[e.g.][]{Akins2021, Karunakaran2021}. Within the FIRE-2 simulations, \citet{Samuel2022} show that a wide range of quenching definitions produce broadly consistent results for low-mass satellites, supporting the use of simple, observationally motivated thresholds.

For the FIRE simulations, we estimate the SFR using an archaeological method by summing the mass of all star particles formed within the past 10 Myr in the particle data extracted for each satellite within its halo region. This yields a 10 Myr averaged SFR that captures star formation on the timescale probed by H$\alpha$ in FIRE-2 \citep{FloresVelazquez2021}. We define the specific star formation rate (sSFR) as
\begin{equation}
\mathrm{sSFR}_{z=0} = \frac{\mathrm{SFR}_{10\,\mathrm{Myr}}}{M_{\star}^{\mathrm{sat}}}.
\end{equation}
In the FIRE simulations, $M_{\star}^{\mathrm{sat}}$ is the bound stellar mass measured within $r^{\star}_{80}$, while the 10 Myr SFR is computed from the satellite particle files extracted within the halo region.
We classify satellites as quenched if $\log_{10}(\mathrm{sSFR}_{z=0}) < -11$, consistent with the approximate division used in \citet{SAGA_IV}. For TNG50, we adopt the instantaneous SFR from the public catalog and apply the same threshold. We verified that our qualitative results remain robust to modest variations in the adopted quenching threshold. In particular, thresholds between $\log_{10}(\mathrm{sSFR}_{z=0}) = -11.5$ and $-10.5$ produce nearly identical quenched fraction trends, while more permissive thresholds systematically increase the overall quenched fractions without changing the relative behavior between simulations or the anomalous radial trend in the paired FIRE-2 systems.

For the SAGA DR3 dataset, we adopt the primary quenching criterion defined in \citet{SAGA_IV}, which classifies confirmed satellites as quenched when
\begin{equation}
{\rm EW}_{\mathrm{H}\alpha} - \sigma_{{\rm EW}_{\mathrm{H}\alpha}} < 2\,\text{\AA},
\end{equation}
where ${\rm EW}_{\mathrm{H}\alpha}$ is the H$\alpha$ equivalent width and $\sigma_{{\rm EW}_{\mathrm{H}\alpha}}$ is the associated measurement uncertainty.

\citet{SAGA_IV} supplement this with a secondary NUV-based sSFR criterion that reclassifies a small number of massive satellites with quenched cores but extended star-forming regions. This adjustment applies only to twelve satellites in their full sample. Because the public DR3 catalog does not include stellar-mass uncertainties needed to reproduce the combined error term in the NUV definition, we do not implement this second step. Given that it affects a very small number of relatively massive satellites, this omission is unlikely to influence our results. For unconfirmed satellites, we adopt the published DR3 quenched flag directly.

For the ELVES sample, we use the quenched flag provided in the compiled catalog from \citet{SAGA_IV}. These classifications are based primarily on optical colors, with supporting UV, H$\alpha$, and morphological information where available, as described in \citet{Karunakaran2023}. Because ELVES does not provide uniform spectroscopic or UV-based SFR measurements for all satellites, we retain the catalog definitions for consistency with the survey methodology.

\section{Results}\label{sec:results}

\subsection{Satellite population comparison}\label{sec:results_sats}

\begin{figure}
    \centering
    \includegraphics[width=\linewidth]{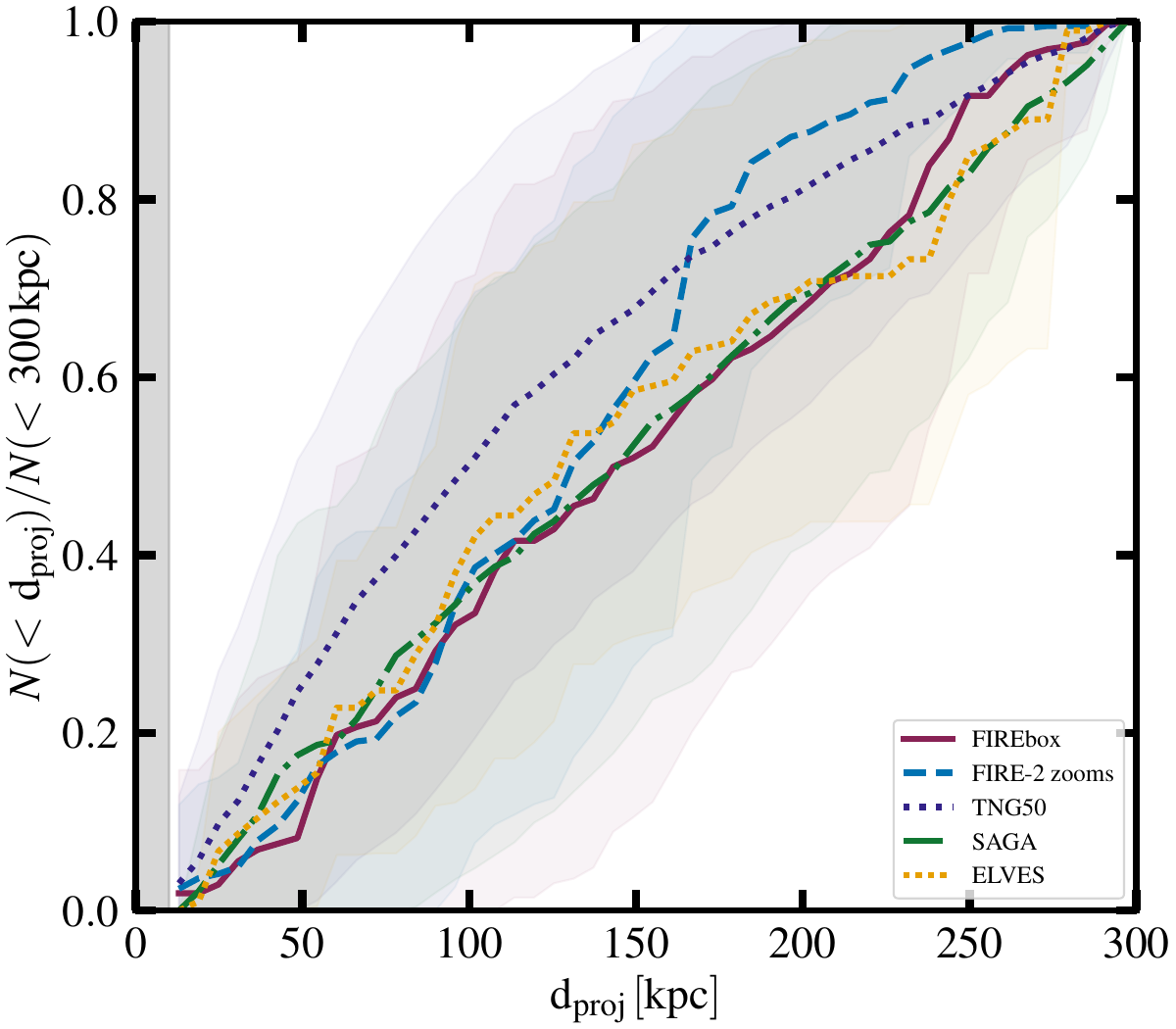}
    \caption{Projected radial distributions of satellites for FIREbox (maroon), the FIRE-2 zoom-ins (blue), TNG50 (indigo), SAGA (green), and ELVES (golden). We plot the median cumulative fraction of satellites per host as a function of projected distance, normalized to the number within 300 kpc. Each simulated host is included along three orthogonal projections. Shaded bands show the $1\sigma$ host-to-host scatter for the simulations. Following the SAGA survey strategy, we exclude the inner 10 kpc (gray region), where host-galaxy light limits satellite detectability. FIREbox closely tracks the observed distributions across the full radial range. The FIRE-2 zoom-ins agree with the observations at small projected distances but become more centrally concentrated at larger radii, while TNG50 is more centrally concentrated across the full radial range.}
    \label{fig:proj_radial_dist}
\end{figure}

In Figure~\ref{fig:proj_radial_dist}, we compare the projected radial distributions of satellites across the three simulation suites and the two observational samples. For every dataset, we compute the cumulative fraction of satellites as a function of projected distance and normalize it to the total number of satellites within 300 kpc. For FIREbox (maroon), the FIRE-2 zoom-ins (blue), TNG50 (indigo), SAGA (green), and ELVES (golden), each curve represents the median cumulative radial profile per host, and the shaded regions for the simulations indicate the $1\sigma$ host-to-host scatter. For the simulations, each host contributes three projected realizations (XY, XZ, and YZ), so the radial distributions reflect all projection-based mock catalogs. As in the SAGA survey, we exclude the innermost 10 kpc (gray shaded region) because strong contamination from the host galaxy light prevents reliable satellite identification at these radii.

The agreement between simulations and observations depends on the simulation suite. FIREbox broadly reproduces the observed radial distributions of SAGA and ELVES closely across the full 10--300 kpc range. The FIRE-2 zoom-ins match the observed satellite distributions well at small projected distances ($\lesssim 150$ kpc), but become more centrally concentrated than the observations at larger radii ($\gtrsim 150$ kpc). TNG50 shows the strongest deviation, with a systematically more centrally concentrated satellite population across the full radial range.

Overall, while not all simulations reproduce the observed radial distributions equally well, FIREbox and the FIRE-2 zoom-ins capture the observed trends within the host-to-host scatter over much of the radial range, whereas TNG50 remains more centrally concentrated than the observations.

\begin{figure}
    \centering
    \includegraphics[width=\linewidth]{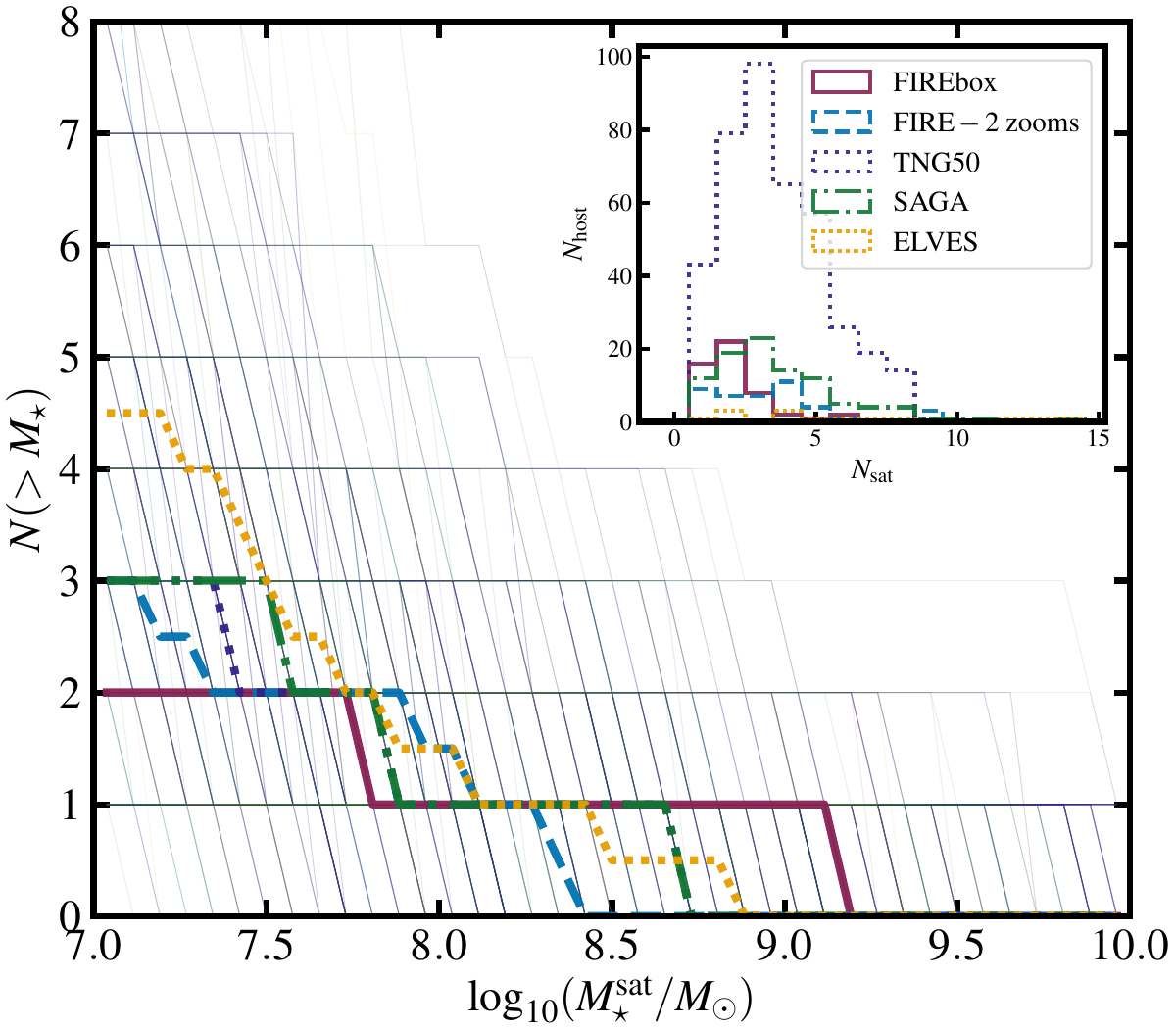}
    \caption{Satellite stellar mass functions (SMFs) for all hosts in FIREbox (maroon), FIRE-2 zoom-ins (blue), TNG50 (indigo), SAGA (green), and ELVES (golden). Thin lines show the SMF for each individual host, while thick lines indicate the median SMF of each dataset. The inset panel shows the distribution of satellite counts per host. The median numbers of satellites in the adopted mass range ($10^{7}$--$10^{10}\,M_\odot$) are 2 for FIREbox, 3 for the FIRE-2 zooms and TNG50, 3 for SAGA, and 4.5 for ELVES.}
    \label{fig:stellar_mass_functions}
\end{figure}

In Figure~\ref{fig:stellar_mass_functions}, we compare the stellar cumulative mass functions (SMFs) of satellites across the full sample of simulated and observed hosts. For the simulations, each thin line represents the SMF of an individual host-projection pair (i.e., each host viewed along the three orthogonal projections), while for the observational datasets each thin line corresponds to a single host. Thick curves represent the median SMF for each dataset, computed over all hosts and projections for the simulations. The inset histogram displays the distribution of the number of satellites per host, providing a complementary view of the abundance statistics.

\begin{figure*}
    \centering
    \includegraphics[width=0.48\textwidth]{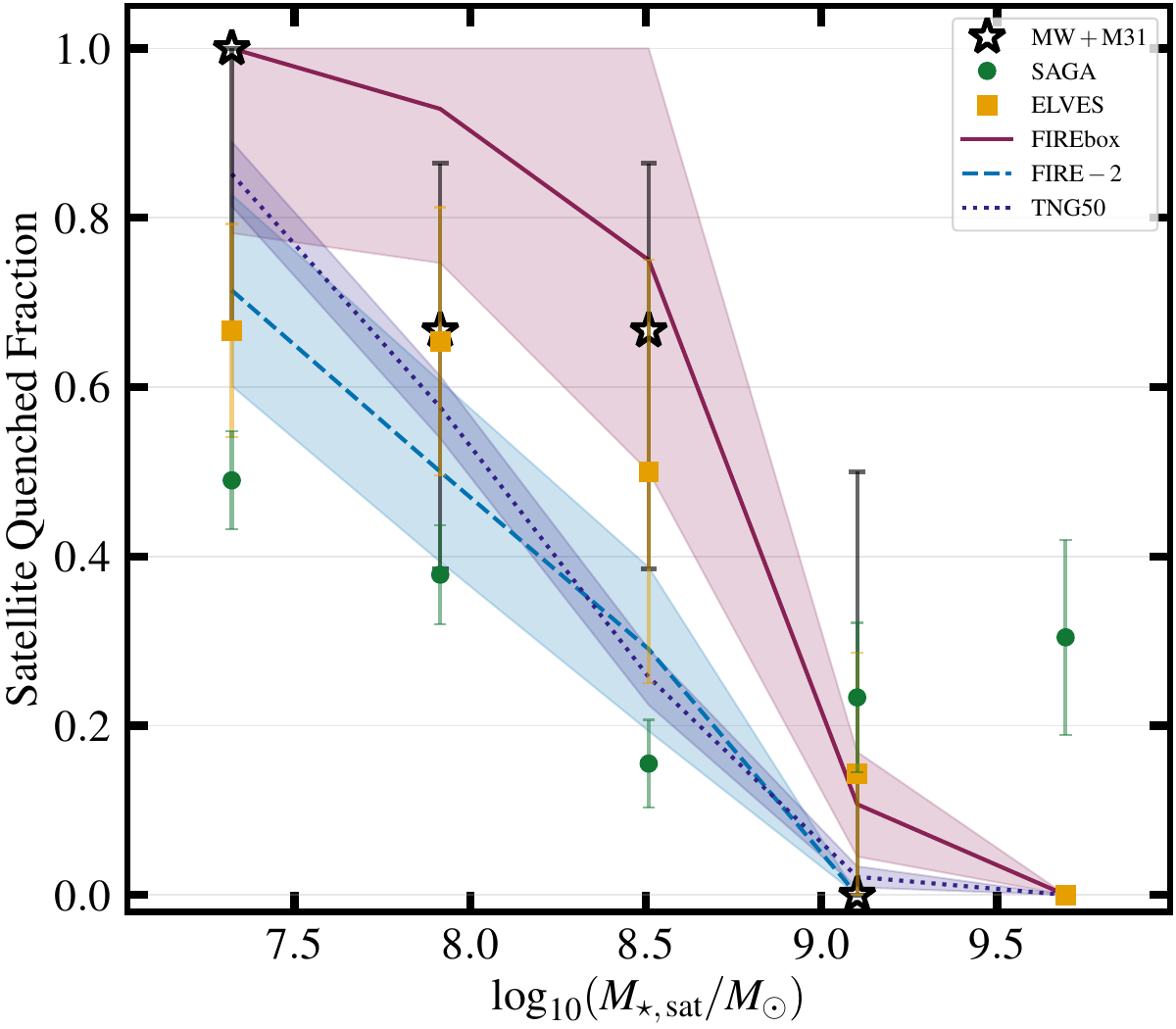}
    \hfill
    \includegraphics[width=0.48\textwidth]{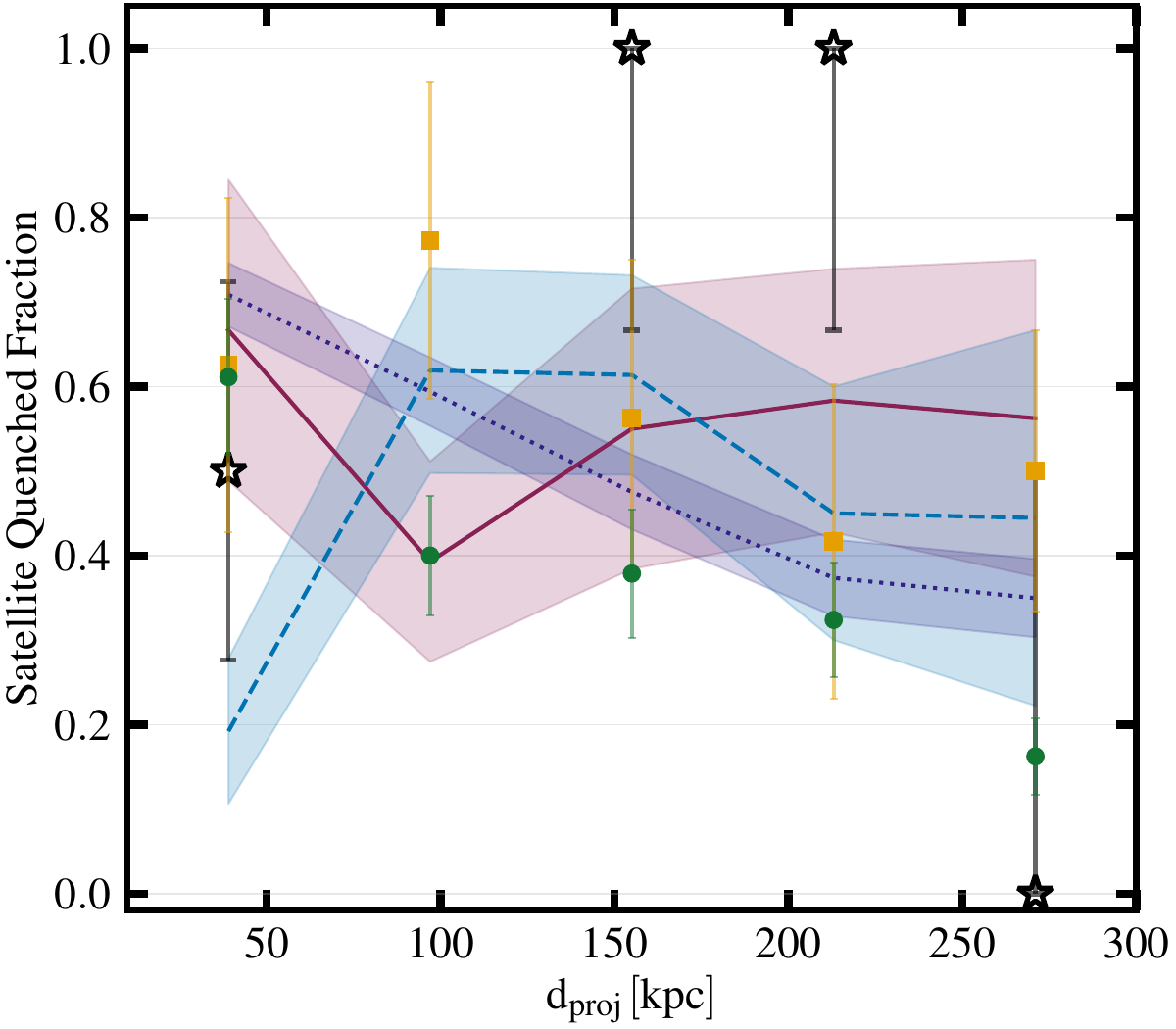}
    \caption{Satellite quenched fractions as a function of stellar mass (left) and projected distance from the host (right). Green circles and golden squares show measurements from the SAGA and ELVES surveys with Poisson uncertainties, and black stars mark the quenched fractions of the combined MW+M31 satellite sample derived from the Local Group dataset of \citet{Wetzel2015}, with error bars showing 68\% binomial confidence intervals. We show simulation predictions from FIREbox (maroon), the FIRE-2 zoom-ins (blue), and TNG50 (indigo) with shaded regions representing Poisson uncertainties on the stacked quenched fractions. The Local Group points use true three-dimensional nearest-host distances and provide a complementary comparison to the projection-based samples. All three simulation suites reproduce the strong rise in quenched fraction toward low stellar masses, in agreement with SAGA, ELVES, and the Local Group hosts. Radial trends show greater diversity: TNG50 most closely follows the observed decline with projected distance, FIREbox remains broadly consistent with a weak radial trend, and the FIRE-2 zoom-ins exhibit suppressed quenched fractions at small projected radii, a behavior that we trace in Section~\ref{sec:results_FIRE_paired} to the paired FIRE-2 host systems. Small-number statistics dominate the Local Group radial measurements, so we use them primarily as a qualitative comparison.}
    \label{fig:quenched_summary}
\end{figure*}

Across all five datasets, the median number of satellites per host remains relatively similar within the stellar-mass range probed here. FIREbox hosts have a median of 2 satellites, whereas the FIRE-2 zoom-ins, TNG50, and SAGA each yield a median of 3. ELVES hosts show a higher median of 4.5 satellites, consistent with their deeper, targeted coverage of the nearby universe. The lower satellite abundance in FIREbox likely reflects a combination of limited volume and sample variance: with only 19 Milky Way–mass hosts drawn from a 22 Mpc box, FIREbox samples fewer environments capable of hosting rich satellite systems, and the intrinsic halo-to-halo scatter at $M_{\rm halo}\sim10^{12}\,M_\odot$ is large. Differences in halo-finding methodology may also contribute, as AHF can be more conservative in identifying heavily stripped subhalos compared to ROCKSTAR or SUBFIND. Despite these variations, the median stellar-mass functions of the three simulation suites agree well with each other and with both observational samples. This consistency indicates that the adopted stellar-mass range ($10^{7}$–$10^{10}\,M_\odot$) yields comparable satellite populations across all datasets and provides a robust foundation for the quenched fraction analysis that follows.

Because our satellite selection is based on projected cylindrical cuts, the resulting samples can include galaxies that are not gravitationally bound to the host. Using the full neighbor catalogs for each simulation, we quantify the fraction of objects within our selection that are true satellites according to the underlying halo catalogs. We find that FIRE-2 zooms and TNG50 yield similar sample purities of $\sim 94\%$ (i.e., $\sim 6\%$ interlopers), while FIREbox exhibits a lower purity of $\sim 68\%$ (i.e., $\sim 32\%$ interlopers).

This difference reflects a combination of halo-finding methodology and modest variations in halo size. FIREbox halos are identified with AHF, which tends to classify nearby objects outside the virial radius as independent centrals, whereas TNG50 (FoF+SUBFIND) and the FIRE-2 zooms (ROCKSTAR) more frequently associate nearby companions with their host. In addition, FIREbox hosts have slightly smaller virial radii, so a larger fraction of nearby galaxies fall outside the formal halo boundary while remaining within the projected aperture. As a result, FIREbox contains a higher fraction of projected interlopers. This highlights that satellite samples defined in projection depend on both halo definitions and aperture choices, an important consideration when comparing to observational surveys such as SAGA and ELVES.

\subsection{Global Satellite Quenched Fractions}\label{sec:results_quenched}

In Figure~\ref{fig:quenched_summary} we present the satellite quenched fractions for FIREbox, the FIRE-2 zoom-in simulations, and TNG50, compared directly with the SAGA and ELVES surveys and with the combined Milky Way and M31 satellite population. For consistency with the observational literature, we compute all quenched fractions in this work, including those from simulations, as stacked quenched fractions. In this definition, the quenched fraction is the total number of quenched satellites divided by the total number of satellites in each bin.\footnote{For each quenched-fraction analysis in this paper, we define stellar-mass bins dynamically using only the datasets included in the corresponding figure. This approach avoids introducing artificial structure from sparsely populated edge bins when comparing different environmental selections or simulation subsets.} For the Local Group comparison, we recompute quenched fractions directly from the satellite counts using the dataset of \citet{Wetzel2015}, applying the same satellite stellar-mass and radial cuts as in our analysis ($7 \leq \log_{10}(M^{\rm sat}_{\star}/M_{\odot}) \leq 10$; $10 \leq {\rm dist/kpc} \leq 300$) and defining quenched galaxies using $M_{\rm gas}/M_\star < 0.1$ (as in \citet{Wetzel2015}). We note that this gas-fraction-based definition differs from the specific star formation rate thresholds typically adopted for SAGA, ELVES, and the simulations, and may introduce systematic offsets in the inferred quenched fractions. We plot Poisson uncertainties for SAGA, ELVES, and all three simulation suites. However, the corresponding host-to-host percentile scatter is typically much larger and often spans most of the allowable range in quenched fraction from 0 to 1. This scatter therefore represents the dominant source of intrinsic variation. For the Local Group points, we instead show 68\% binomial confidence intervals appropriate for the small number of satellites in each bin.

In the left panel we show the quenched fraction as a function of satellite stellar mass. The SAGA and ELVES datasets both exhibit a strong increase in quenched fraction toward lower stellar masses. After applying the same stellar-mass and radial cuts, the Local Group sample, though limited to nine satellites, follows the same qualitative trend, with higher quenched fractions at lower stellar masses. All three simulation suites reproduce this behavior and show qualitative agreement with the observational data. FIREbox predicts somewhat higher quenched fractions over part of the mass range, but these differences remain within the often substantial Poisson uncertainties. TNG50 and the FIRE-2 zoom-ins more closely track the SAGA, ELVES, and Local Group measurements. These results indicate that the stellar-mass dependence of satellite quenching is a robust and well-captured feature across both observations and simulations.

In the right panel, we show the quenched fraction as a function of projected distance from the host. The SAGA and ELVES datasets exhibit a gently declining quenched fraction with radius. Environmental quenching is strongest inside $\sim$50–100 kpc and weakens outward. Among the simulations, TNG50 best reproduces this gradual decline. FIREbox displays a nearly flat trend that remains broadly consistent with the observations once uncertainties are considered. The FIRE-2 zoom-ins depart more noticeably at small radii and show a suppressed quenched fraction inside $\sim 50$ kpc. As shown in Sections~\ref{sec:results_environment} and~\ref{sec:results_FIRE_paired}, this behavior is strongest in the paired FIRE-2 systems, whose atypical satellite populations are associated with much of the unexpectedly low central quenched fraction. The Local Group measurements do not exhibit a smooth radial trend, with individual bins often containing only one or two satellites. The resulting quenched fractions span the full range from 0 to 1, reflecting the sensitivity of this comparison to small-number statistics rather than a well-constrained radial dependence.

Overall, the stellar-mass dependence of satellite quenching shows strikingly good agreement across all three simulation suites and all three observational samples, including the Milky Way and M31. This agreement indicates that the trend is a robust outcome of contemporary galaxy formation models despite their differing volumes, physics implementations, and numerical resolutions. The radial trends, by contrast, are more sensitive to the detailed distribution of satellite stellar masses, radii, and host-to-host variations.

\subsection{Environmental dependence of satellite quenching}
\label{sec:results_environment}

\begin{figure*}
    \centering
    \includegraphics[width=\textwidth]{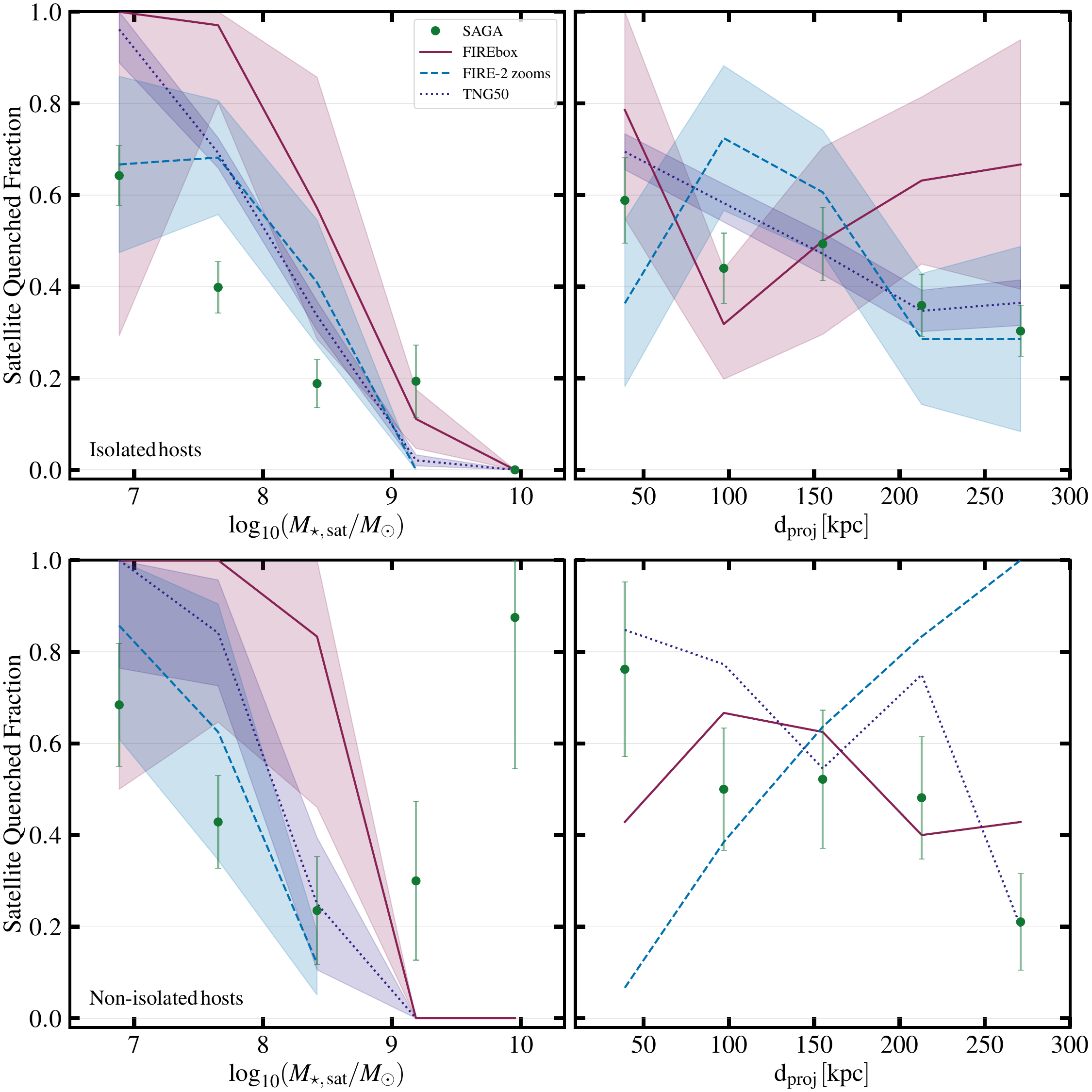}
    \caption{Satellite quenched fractions split by host environment. The top row shows isolated hosts, and the bottom row shows non-isolated hosts. Left panels show quenched fraction as a function of satellite stellar mass, while right panels show quenched fraction as a function of projected host-centric distance. Green circles show SAGA, with Poisson uncertainties. We show simulation predictions from FIREbox (maroon), the FIRE-2 zoom-ins (blue), and TNG50 (indigo) as lines, with shaded regions indicating Poisson uncertainties for all panels except the non-isolated radial panel, where we omit simulation uncertainties for visual clarity. For isolated hosts, all three simulations broadly reproduce the observed stellar-mass trends and show generally flat or declining radial quenched fractions, although the FIRE-2 zoom-ins exhibit a modest suppression of quenched satellites at small radii. For non-isolated hosts, FIREbox and TNG50 remain broadly consistent with a declining radial quenched fraction, while the FIRE-2 zoom-ins show a strongly inverted radial trend, with quenched fraction increasing toward larger projected distance. This behavior drives the suppressed inner quenched fraction seen in the global FIRE-2 radial trend in Figure~\ref{fig:quenched_summary}, and we examine it in detail in Section~\ref{sec:results_FIRE_paired}.}
    \label{fig:qf_environment}
\end{figure*}

We next examine whether the quenched fraction trends depend on host environment by dividing each sample into isolated and non-isolated hosts, following the definitions in Section~\ref{sec:environment_definition}. Figure~\ref{fig:qf_environment} shows quenched fractions as a function of satellite stellar mass and projected distance for the two environment classes.

For isolated hosts (top row), all three simulation suites reproduce the observed stellar-mass trend, with quenched fractions increasing toward lower satellite stellar mass. As in the full sample, the simulations tend to overpredict the quenched fraction relative to SAGA at $M_\star < 10^{9}\,M_\odot$. The radial trends are noisier, particularly for FIREbox and the FIRE-2 zoom-ins, reflecting small satellite numbers per host. FIREbox and TNG50 are broadly consistent with a flat or gradually declining quenched fraction with projected distance. The isolated FIRE-2 zoom-in hosts follow a similar overall trend, but still show a noticeable deficit of quenched satellites at small projected distances ($\lesssim 100$ kpc), indicating that the inner suppression is present even in the absence of nearby massive companions, albeit at a reduced level.

For non-isolated hosts (bottom row), the stellar-mass trends remain broadly similar across datasets at low and intermediate satellite masses. SAGA exhibits a sharp upturn in quenched fraction in the highest stellar-mass bin, $M_\star \sim 10^{9}$--$10^{10}\,M_\odot$, which lies near our imposed upper mass cut and is not well populated in the simulated samples.

In projected distance, FIREbox and TNG50 again show radial trends consistent with SAGA, with quenched fractions that are roughly flat or decline toward larger radii. In contrast, the FIRE-2 zoom-ins display a qualitatively different behavior: their non-isolated hosts exhibit a clear monotonic increase in quenched fraction with projected distance, producing a strongly inverted radial relation. Compared to the isolated FIRE-2 hosts, where the inner suppression is present but modest, this inversion is significantly amplified in the non-isolated systems. These results suggest that the anomalous radial trend in the FIRE-2 zoom-ins is not driven solely by environment, but is instead associated with the particular satellite population sampled by the paired MW-M31 analogs. We investigate the origin of this behavior in the following subsection.

\subsection{Discrepancy in the FIRE-2 zoom-ins}
\label{sec:results_FIRE_paired}

\begin{figure*}
    \centering
    \includegraphics[width=\textwidth]{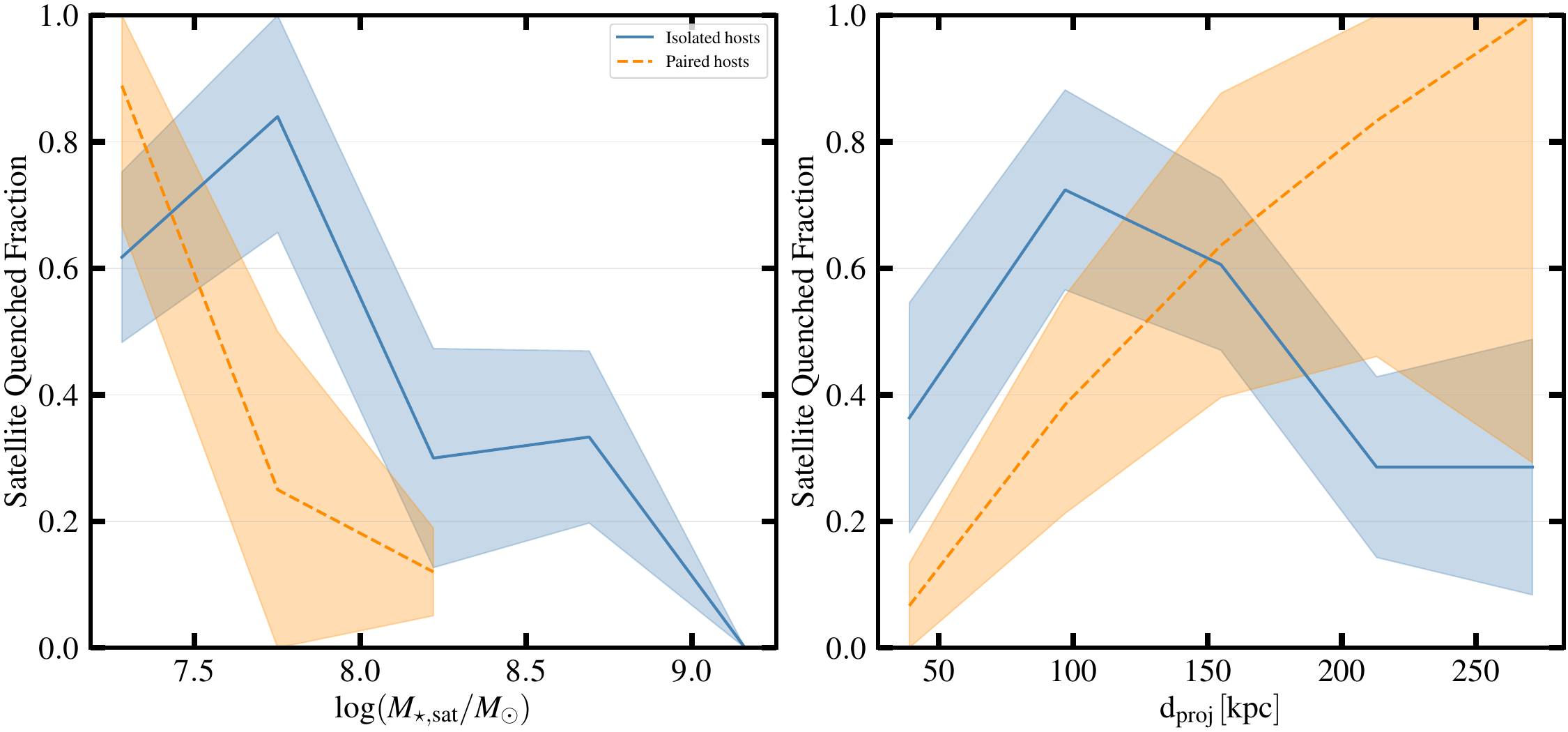}
    \caption{Satellite quenched fractions in the FIRE-2 zoom-ins split by host environment: isolated MW--mass analogs (blue) and paired MW-M31 analogs (orange). \textit{Left:} As a function of satellite stellar mass, the paired hosts exhibit a distinct satellite stellar-mass distribution compared to the isolated systems, with no satellites above $M_\star \approx 10^{8.5}\,M_\odot$. \textit{Right:} Satellites of paired hosts show a rising quenched fraction with increasing $d_{\rm proj}$, whereas the isolated FIRE-2 hosts exhibit a noisier but non-monotonic radial trend with only a modest suppression of quenched satellites at small projected distances. The paired systems therefore drive the discrepant behavior seen in the full FIRE-2 sample.}
    \label{fig:qf_iso_pairs}
\end{figure*}

The FIRE-2 zoom-ins display the strongest discrepancy among the simulations in their radial quenched fraction trends (Section~\ref{sec:results_quenched}). To investigate the origin of this behavior, we divide the FIRE-2 hosts into isolated Milky Way analogs and paired MW-M31 analogs. We emphasize that we use this division as a diagnostic to isolate the source of the discrepancy, rather than to attribute it directly to environmental effects. In Figure~\ref{fig:qf_iso_pairs} we compare the quenched fractions for satellites around these two subsets as a function of both satellite stellar mass (left) and projected host-centric distance (right).

In the stellar-mass panel, paired systems (orange) lack satellites above $M_\star \approx 10^{8.5}\,M_\odot$, whereas isolated hosts (blue) contain satellites extending to $M_\star \gtrsim 10^{9}\,M_\odot$. \citet{Samuel2022} present a similar comparison in their Fig.~5 and likewise report that the paired systems host fewer massive satellites, although their exact trends differ because they use a stricter quenching threshold and include satellites down to lower stellar masses. This difference highlights that the paired FIRE-2 systems sample a distinct satellite population relative to the isolated systems. However, the absence of high-mass satellites alone does not fully explain the discrepant radial quenched-fraction trend, since other non-isolated host samples also exhibit relatively few high-mass satellites without producing the same inverted radial behavior (see bottom row of Figure~\ref{fig:qf_environment}).

We verified that this discrepancy persists when restricting the analysis to low-mass satellites with $M_\star < 10^{9}\,M_\odot$. In the global radial quenched-fraction comparison, FIREbox and TNG50 remain broadly consistent with the observations, showing flat or gently declining radial trends, while the FIRE-2 zoom-ins continue to show suppressed quenched fractions at small projected distances. Similarly, when applying this low-mass cut to the environment-split analysis, the non-isolated FIREbox and TNG50 hosts do not develop inverted radial trends. These tests confirm that the FIRE-2 behavior is not driven solely by the absence of high-mass satellites in the paired systems.

The radial trends show an even sharper contrast. The isolated hosts exhibit a noisier but broadly conventional radial behavior, although they still show somewhat suppressed quenched fractions within $\sim 100$ kpc. Satellites of the paired hosts, however, show the opposite behavior: the quenched fraction is lowest at small projected radii and increases toward larger distances. This inverted trend directly reproduces the discrepancy highlighted in Figure~\ref{fig:quenched_summary} and demonstrates that the paired systems dominate the anomalous radial behavior in the FIRE-2 suite. Importantly, this inversion does not by itself imply a fundamentally different quenching mechanism.

\begin{figure*}
    \centering
    \includegraphics[width=\textwidth]{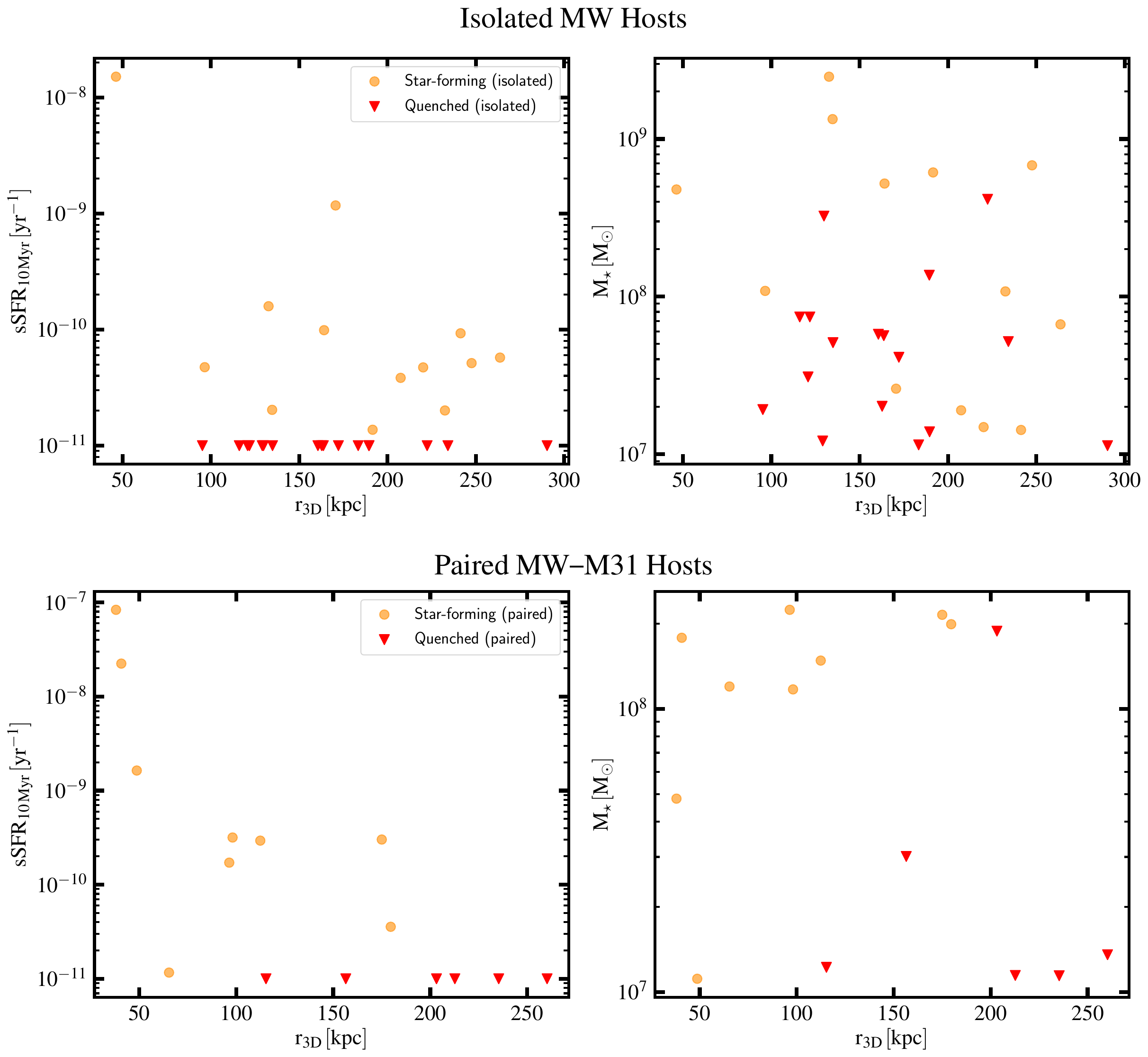}
    \caption{Structural properties of satellites in the FIRE-2 zoom-ins, separated by host environment. \textit{Top:} Satellites of the isolated MW analogs. Star-forming (orange circles) and quenched (red triangles) satellites span a wide range of radii and stellar masses, including systems as massive as $M_\star \sim 3\times10^{9}\,M_\odot$, and extend across the full radial range.
    \textit{Bottom:} Satellites of the paired MW-M31 analogs. Star-forming satellites are concentrated at small radii and higher stellar masses, while quenched satellites occupy larger radii and lower stellar masses. 
    Quenched satellites are found primarily beyond $\sim 100$ kpc and near $M_\star \sim 10^{7}\,M_\odot$, while more massive satellites are preferentially star-forming and located at smaller radii. These structural differences underlie the divergent quenched fraction trends seen in Figure~\ref{fig:qf_iso_pairs}.}
    \label{fig:sf_mass_distance_panels}
\end{figure*}

To further explore this behavior, we examine satellite properties as a function of their 3D host-centric distances in Figure~\ref{fig:sf_mass_distance_panels}. We display true satellites around isolated hosts in the top row and those around paired hosts in the bottom row. Star-forming satellites appear as orange circles and quenched satellites as red triangles; we adopt $\rm sSFR_{10,Myr} = 10^{-11}\,{\rm yr}^{-1}$ for all quenched satellites.

Two trends stand out. First, satellites of the isolated hosts span a broad range in both stellar mass and radius, forming a relatively extended and mixed population across the full radial range. Second, the paired hosts exhibit strong segregation between star-forming and quenched satellites. Star-forming satellites are preferentially concentrated at smaller radii, while quenched satellites are found primarily at larger distances from the host. In particular, quenched satellites in the paired systems are largely confined to $r_{3D} \gtrsim 100$ kpc, whereas satellites at smaller radii are predominantly star-forming.

These results suggest that the inverted quenched-fraction trend in the paired FIRE-2 systems reflects an atypical satellite population with strong spatial segregation between star-forming and quenched satellites. These differences likely reflect variations in assembly history and the limited sampling of paired systems, rather than a universal effect of large-scale environment on satellite quenching. A definitive interpretation will require an analysis of satellite orbits, infall times, and gas histories, which we defer to future work.

\section{Discussion}\label{sec:discussion}

\subsection{Implications of our results}\label{sec:discussion_implications}

Comparing satellite populations across simulations presents several challenges, even when host mass ranges and selection criteria are carefully matched. Differences in halo-finding algorithms, numerical resolution, and feedback implementations each introduce systematic variations that complicate direct comparisons. These differences affect both the inferred properties of halos and the resulting star formation histories of their satellites.

Despite these challenges, our results show that the \emph{mass dependence} of satellite quenching is a robust prediction of contemporary galaxy-formation models. All three simulation suites reproduce the strong trend that lower-mass satellites are more likely to be quenched (Figure~\ref{fig:quenched_summary}, left panel), in excellent qualitative agreement with observations. This consistency suggests that the essential coupling between stellar feedback, gas removal, and environmental processing within host halos is broadly captured across very different numerical frameworks.

By contrast, the \emph{radial} quenched-fraction trends are more sensitive to the detailed satellite populations being sampled. Here the FIRE-2 zoom-ins diverge most strongly from the observations and the other simulations. We show in Sections~\ref{sec:results_environment} and~\ref{sec:results_FIRE_paired} that this behavior is strongest in the paired FIRE-2 systems, whose satellite populations differ markedly from their isolated counterparts. In particular, the paired hosts exhibit strong spatial segregation between star-forming and quenched satellites, producing an inverted quenched-fraction profile in which the lowest quenched fractions occur at the smallest projected radii.

We do not interpret this behavior as evidence that large-scale environment directly sets the quenched fraction of satellites. Instead, the FIRE-2 discrepancy appears to reflect an atypical satellite population in the paired systems, characterized by a distinct spatial and star-forming distribution of satellites. Understanding why those satellites occupy this configuration requires a more detailed analysis of their orbits, infall times, and gas histories. Thus, our results indicate that radial quenched-fraction trends can be strongly affected by host-to-host variations in satellite populations, even when the stellar-mass dependence of quenching remains robust.

Host properties themselves likely play an important role in shaping satellite quenching. The SAGA survey shows that the lowest-mass satellites around the most massive hosts exhibit systematically higher quenched fractions, particularly at small projected distances from the host, suggesting that host mass can influence the efficiency of satellite processing and quenching \citep{SAGA_IV}. \citet{MartinNavarro2021} further argue that AGN activity can potentially enhance quenching anisotropically, preferentially affecting satellites along the axis of a black hole jet. However, recent work by \citet{Karp2023} demonstrates that such anisotropic quenching is unlikely to be driven by AGN feedback. Instead, their analysis of TNG50 finds strong evidence that the signal originates from anisotropy in satellite accretion times, with satellites accreted earlier experiencing more prolonged environmental processing. These results suggest that anisotropic satellite quenching is more naturally explained by the geometry and timing of satellite accretion than by AGN-driven feedback. Such findings reinforce the need to connect quenched fractions to satellite orbital histories before attributing differences to a single physical mechanism.

It is also important to consider how representative the simulation suites are of the broader galaxy population. FIREbox, while high resolution, spans a modest cosmological volume and may under-sample rare environments. The FIRE-2 zoom-ins are selected to be MW-mass halos but are not tailored to reproduce detailed Milky Way or Local Group satellite populations; several of the zoom-ins lack massive satellites such as the LMC or M33. These limitations reinforce the need to interpret radial quenched-fraction trends in the context of the particular satellite populations present in each host sample.

\subsection{Additional factors not explored in this work}\label{sec:discussion_additional}

Several physical ingredients relevant to satellite quenching lie beyond the scope of the present analysis. One is the detailed gas content of satellites. Previous work shows that most satellites within the virial radius of MW-like hosts are gas poor, having been stripped of their gas after infall, while the low-mass galaxies that retain significant gas are generally isolated systems \citep{Simpson2018, Samuel2022, Engler2023}. \citet{Akins2021} find that quenching timescales correlate positively with gas mass at infall, suggesting that gas-rich satellites can remain star-forming long after entering a host halo.

Projection effects and interlopers also influence inferred quenched fractions. In this work, we partially address these effects by constructing projection-based mock catalogs and quantifying interloper fractions in the simulations. However, a fully survey-realistic treatment including observational incompleteness and surface-brightness limits remains beyond the scope of the present analysis.

Work at other host mass scales also provides broader context. Observational and simulation studies explore quenched fractions around lower-mass hosts \citep[e.g.,][]{Jahn2022, Li2025} and around group- and cluster-scale systems \citep[e.g.,][]{Donnari2021, Baxter2021}. These studies show that the efficiency and dominant mechanisms of quenching vary significantly with host mass, suggesting that satellite quenching reflects a continuum of processes operating across halo scales. Integrating insights across these regimes will be important for constructing a unified view of low-mass galaxy quenching.

\subsection{Future directions}\label{sec:discussion_future}

The analysis presented here points to two important directions for future work. First, the FIRE-2 zoom-in discrepancy should be investigated through the orbital and gas-accretion histories of individual satellites. The inverted radial quenched-fraction trend in the paired systems is associated with an unusual spatial distribution of star-forming and quenched satellites but a definitive interpretation requires infall times, pericentric passages, preprocessing histories, and gas-loss timescales. A detailed analysis of satellite orbits, infall times, and gas histories in the FIRE-2 zoom-ins (Gogliettino \& Gandhi et al., in prep) will directly address this question.

Second, improved mock-observational comparisons are needed to determine how projection effects, interlopers, and survey selection functions influence measured quenched fractions. A study using FIREbox together with a subset of FIRE-2 zoom-ins (Tortora et al., in prep) will construct full mock observations to evaluate selection effects and assess how faithfully simulated satellites can be compared to systems identified in surveys such as SAGA and ELVES.

Together, these efforts will clarify why radial quenched-fraction trends vary more strongly across datasets than stellar-mass trends, and will help connect satellite quenching to the detailed assembly histories of Milky Way--mass systems.

\section{Summary and Conclusions}\label{sec:conclusion}

In this work, we compare satellite populations across three state-of-the-art cosmological simulation suites (FIREbox, the FIRE-2 zoom-ins, and TNG50), two large observational datasets (SAGA and ELVES), and the combined Milky Way and M31 satellite population. By applying nearly consistent host and satellite selection criteria, along with common quenching definitions, we examine how well contemporary galaxy-formation models reproduce the observed dependencies of satellite quenching on stellar mass and projected host distance.

Our main conclusions are as follows:

\begin{itemize}
    \item \textbf{Mass dependence of quenching is robust.}  
    All three simulation suites reproduce the strong empirical trend that lower-mass satellites are more likely to be quenched. This agreement extends across the SAGA and ELVES surveys and the combined MW+M31 sample, indicating that the simulations capture the same mass-dependent behavior seen in the full set of available observations. Despite substantial differences in numerical resolution, halo-finding methodology, and feedback modeling, the consistency across both simulations and observations demonstrates that the mass dependence of satellite quenching is a robust outcome of contemporary galaxy-formation models.
    
    \item \textbf{Radial quenched fraction trends broadly agree, but are more sensitive to satellite population details.}
    Observations from SAGA and ELVES show gently declining quenched fractions with projected distance, reflecting stronger environmental quenching at smaller radii. TNG50 most closely reproduces this decline, while FIREbox remains broadly consistent with only weak radial variation across the probed range. The FIRE-2 zoom-ins show suppressed quenched fractions at small projected distances, particularly in the paired MW-M31 analogs. Unlike the stellar-mass trends, which remain remarkably stable across all datasets and environments, the radial quenched fractions are substantially more sensitive to the detailed stellar-mass and spatial distributions of the underlying satellite populations.

    \item \textbf{The discrepancy in FIRE-2 is amplified in the paired MW-M31 analogs, but is not driven solely by environment.}
    Splitting the FIRE-2 hosts into isolated and paired systems reveals that the strongest inversion of the radial quenched-fraction trend occurs in the paired hosts, where quenched fractions increase with projected distance. However, the isolated FIRE-2 hosts also exhibit a modest deficit of quenched satellites at small radii relative to the observations and the other simulations. The paired systems lack high-mass satellites and show strong radial segregation between star-forming and quenched satellites, but the absence of high-mass satellites alone does not explain the radial discrepancy. Instead, the behavior appears linked to the atypical joint distribution of satellite stellar mass, radius, and quenching state in these systems. Understanding the origin of this behavior will require detailed analyses of satellite orbits, infall times, and gas histories.

    \item \textbf{Host environment does not strongly alter the stellar-mass dependence of quenching.}  
    Across both isolated and non-isolated hosts, the simulations and observations show qualitatively similar quenched fraction trends with satellite stellar mass. This indicates that the stellar-mass dependence of satellite quenching is comparatively insensitive to large-scale host environment over the range explored in this work. Differences between environment classes are much more apparent in the radial trends, where the detailed spatial and stellar-mass distributions of satellites become increasingly important.

    \item \textbf{Satellite populations are broadly similar across datasets in number and stellar mass.}  
    Across FIREbox, FIRE-2, TNG50, SAGA, and ELVES, the satellite stellar mass functions and projected radial distributions are broadly comparable, although systematic differences remain between datasets. In particular, TNG50 exhibits a more centrally concentrated satellite population than is observed in SAGA and ELVES. Median satellite counts per host differ only modestly across datasets, supporting the validity of comparing quenched fractions within a unified stellar-mass range of $10^{7}$-$10^{10}\,M_\odot$.

    \item \textbf{Interpretation requires caution due to observational and numerical systematics.}
    Differences in halo finders, simulation volumes, resolution limits, and observational selection functions each introduce systematic uncertainties. In this work, we partially address these effects through projection-based mock catalogs and explicit interloper measurements, but fully survey-realistic analyses will still be important for future quantitative comparisons.
\end{itemize}

Our results demonstrate that contemporary simulations successfully reproduce the strong stellar-mass dependence of satellite quenching across a wide range of datasets and numerical methods. By contrast, radial quenched-fraction trends are more sensitive to the detailed structure of the satellite populations within individual host systems. The behavior of the FIRE-2 paired systems illustrates how variations in satellite stellar-mass distributions and radial segregation can strongly modify inferred radial quenched fractions without necessarily implying a fundamentally different quenching mechanism. These findings highlight the importance of connecting satellite quenching statistics to the detailed assembly histories and orbital properties of satellite populations.

Future work combining orbital histories, gas evolution, and improved mock-observational analyses will be essential for determining how satellite populations acquire the structural properties that ultimately shape their observed quenched-fraction trends.


\section*{Acknowledgments}
FJM is funded by the National Science Foundation (NSF) Math and Physical Sciences (MPS) Award AST-2316748. DCB is supported by an NSF Astronomy and Astrophysics Postdoctoral Fellowship under award AST-2303800 as well as by the Chancellor's Postdoctoral Fellowship. MKRW acknowledges support from NSF MPS Ascending Faculty Catalyst Award AST-2444751. JM is funded by a Pomona College Large Research Grant. AW received support from NSF, via CAREER award AST-2045928 and grant AST-2107772.

We thank the anonymous referee for insightful comments and suggestions that greatly improved the quality and clarity of this manuscript. We also thank Marla Geha for helpful discussions and suggestions that improved the interpretation of our results.

This research used data from the SAGA Survey (Satellites Around Galactic Analogs; sagasurvey.org). The SAGA Survey is a galaxy redshift survey with spectroscopic data obtained by the SAGA Survey team with the Anglo-Australian Telescope, MMT Observatory, Palomar Observatory, W. M. Keck Observatory, and the South African Astronomical Observatory (SAAO). The SAGA Survey also made use of many public data sets, including: imaging data from the Sloan Digital Sky Survey (SDSS), the Dark Energy Survey (DES), the GALEX Survey, and the Dark Energy Spectroscopic Instrument (DESI) Legacy Imaging Surveys, which includes the Dark Energy Camera Legacy Survey (DECaLS), the Beijing-Arizona Sky Survey (BASS), and the Mayall z-band Legacy Survey (MzLS); redshift catalogs from SDSS, DESI, the Galaxy And Mass Assembly (GAMA) Survey, the Prism Multi-object Survey (PRIMUS), the VIMOS Public Extragalactic Redshift Survey (VIPERS), the WiggleZ Dark Energy Survey (WiggleZ), the 2dF Galaxy Redshift Survey (2dFGRS), the HectoMAP Redshift Survey, the HETDEX Source Catalog, the 6dF Galaxy Survey (6dFGS), the Hectospec Cluster Survey (HeCS), the Australian Dark Energy Survey (OzDES), the 2-degree Field Lensing Survey (2dFLenS), and the Las Campanas Redshift Survey (LCRS); HI data from the Arecibo Legacy Fast ALFA Survey (ALFALFA), the FAST all sky HI Survey (FASHI), and HI Parkes All-Sky Survey (HIPASS); and compiled data from the NASA-Sloan Atlas (NSA), the Siena Galaxy Atlas (SGA), the HyperLeda database, and the Extragalactic Distance Database (EDD). The SAGA Survey was supported in part by NSF collaborative grants AST-1517148 and AST-1517422 and Heising–Simons Foundation grant 2019-1402. SAGA Survey's full acknowledgments can be found at https://sagasurvey.org/ack/.


\software{The functionalities provided by the following python packages played a critical role in the analysis and visualizations presented in this paper: \textbf{\textsc{matplotlib}} \citep{Hunter2007}, \textbf{\textsc{NumPy}} \citep{vanderWalt2011}, \textbf{\textsc{SciPy}} \citep{Virtanen2020} and \textbf{\textsc{iPython}} \citep{Perez2007}.}


\bibliography{refs}{}
\bibliographystyle{aasjournal}

\end{document}